\documentclass[10pt,titlepage,letterpaper,twoside,twocolumn]{article}

\usepackage{amsmath}
\usepackage{graphicx} 
\usepackage[compress,numbers]{natbib}
\title{MoNA - The first 25 years}
\date{January 2025}

\begin{document}

\author{A. Banu}
\author{T.~Baumann}
\author{J.~Brown}
\author{P.~A.~DeYoung}
\author{J.~Finck}
\author{N.~Frank}
\author{P.~Gu{\`e}ye}
\author{J.~Hinnefeld}
\author{C.~Hoffman}
\author{A.~N.~Kuchera}
\author{B.~A.~Luther}
\author{B.~Monteagudo-Gudoy}
\author{T.~Redpath}
\author{A.~Revel}
\author{W.~F.~Rogers}
\author{M. Thoennessen}

\onecolumn
\pagestyle{empty}
\begin{titlepage}	
  \vspace*{2in}%
	\parbox{0.4\textwidth}{%
  \includegraphics[width=0.3\textwidth]{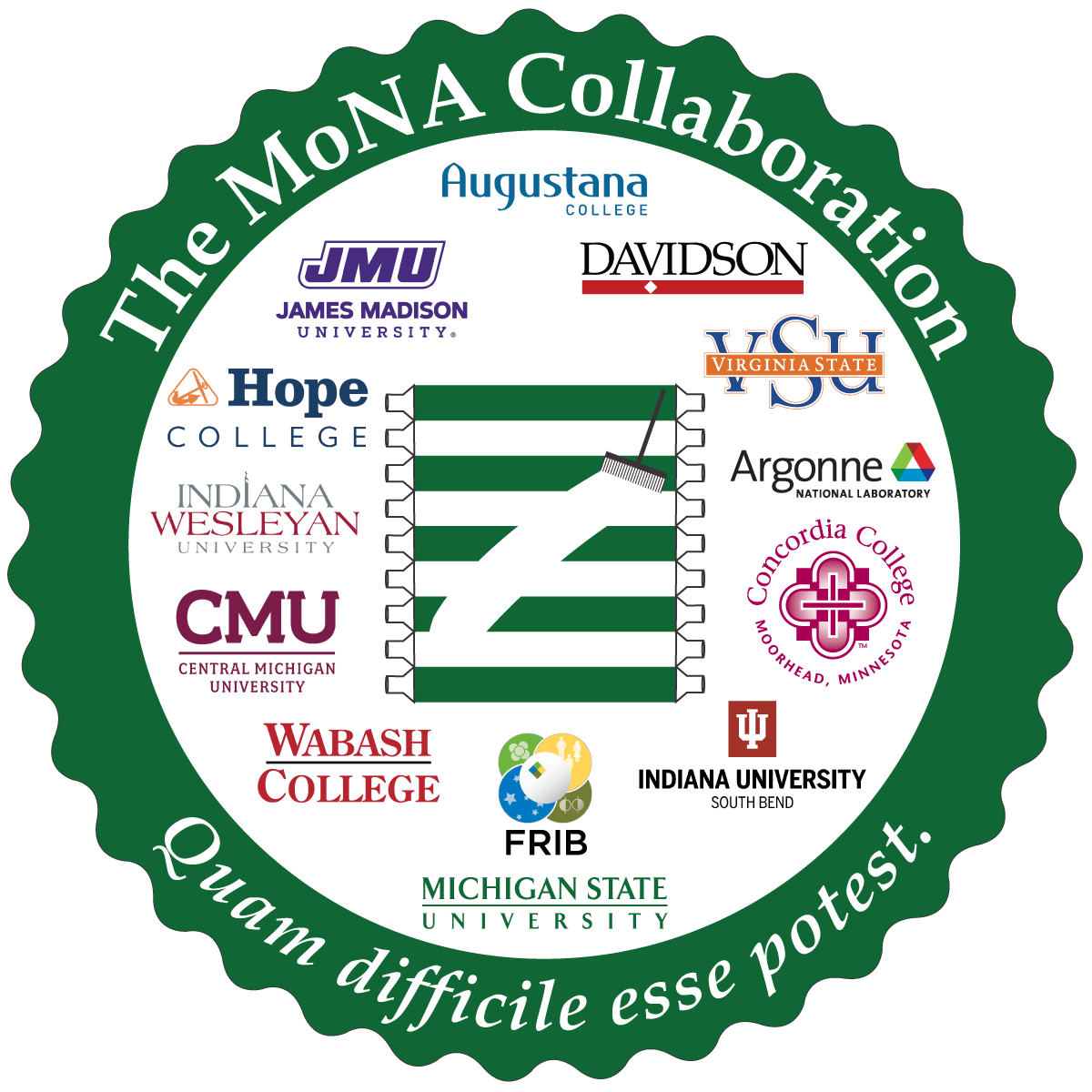}
	}%
  \parbox{0.7\textwidth}{%
    {\huge {\bfseries MoNA} - The first 25 years}%
    \vskip 5ex%
    {\large\itshape The MoNA Collaboration}\\[2ex]
A. Banu, T.~Baumann, J.~Brown, P.~A.~DeYoung, \\
J.~Finck, N.~Frank, P.~Gu{\`e}ye, J.~Hinnefeld, \\
C.~Hoffman, A.~N.~Kuchera, B.~A.~Luther, \\
B.~Monteagudo-Gudoy, T.~Redpath, A.~Revel, \\
W.~F.~Rogers, M. Thoennessen\\[5ex] 

\textit\footnotesize{%
Argonne National Laboratory, Lemont, IL 60439\\
Augustana College, Rock Island, IL 61201\\
Central Michigan University, Mount Pleasant, MI 48859\\
Concordia College, Moorhead, MN 56562\\
Davidson College, Davidson, NC 28035\\
James Madison University, Harrisonburg, VA 22807\\
Hope College, Holland, MI 49423\\
Indiana University South Bend, South Bend, IN 46634\\
Indiana Wesleyan University, Marion, IN 46953\\
Michigan State University, East Lansing, MI 48824-1321\\
Virginia State University, St. Petersburg, VA 23806\\
Wabash College, Crawfordsville, IN 47933}
    \vskip 8ex%
    \today%
    }
\end{titlepage}

\clearpage

\twocolumn

\section{Introduction}
The last decade of the twentieth century saw a resurgence in nuclear structure physics. Intense secondary beams of radioactive nuclei became available that substantially expanded the horizon of the nuclear chart, especially near the neutron-rich side. The National Superconducting Cyclotron Laboratory (NSCL) at Michigan State University (MSU) was one of the few laboratories in the world that had developed these capabilities. NSCL researchers had begun an extensive program to study exotic nuclei at and beyond the neutron dripline, for example, the Borromean nucleus $^{10}$Li \cite{Kryger10Li} and the halo structure of $^{11}$Li \cite{Orr11Li}. 

In 1996, a project was initiated at the NSCL to couple the two cyclotrons, including the construction of a larger-acceptance fragment separator. This new Coupled Cyclotron Facility (CCF) \cite{CCF} offered significantly higher secondary beam intensities. It became clear that an efficient neutron detector would be needed to improve the study of interesting two-neutron halo systems, excited states of near-dripline nuclei, and unbound resonances just beyond the dripline. Such a device would have to be position sensitive, have multi-hit capability, and have good time resolution suitable for time-of-flight methods. These considerations pushed towards a modular design, a feature that fortuitously led to the formation of the Modular Neutron Array (MoNA) collaboration. The very physics and budgetary constraints that had driven for a modular design also allowed for a simple collaborative solution of partitioning the work of detector development across multiple institutions, with the added benefit of allowing undergraduate students to experience and be involved in world-class research. 

Over the last 25 years, the collaboration was carried by the combined efforts of hundreds of undergraduates, dozens of graduate students and research associates, and faculty from more than a dozen colleges and universities, produced over fifty publications, won awards for research, and combined research and teaching in new and interesting ways. The original group of MoNA institutions included many different styles of higher education institutions, including liberal arts colleges, regional comprehensive universities, and major research universities. That diversity of styles is still reflected in the members to this date. 

The past achievements of the collaboration include the discovery of seven of the sixteen currently known unbound isotopes, from lithium to fluorine. The construction of the Facility for Rare Isotope Beams (FRIB) \cite{FRIB} at MSU has opened opportunities for further and more detailed studies. This new facility requires a new approach and the current efforts of the collaboration to build the Next Generation neutron detector (NGn) reflect this. Recognizing a trade-off of position resolution and efficiency, the new array will emphasize resolution reflecting the higher level densities expected as we move towards the heavier nuclei accessible by FRIB. While the collaboration looks forward to these new studies, in the following sections we look back over the last quarter century from the ideation and building of MoNA, through LISA, to the threshold of NGn, explore the scientific impacts of the experiments enabled by these devices and the educational impact that comes through the shared work of the growing collaboration.

\section{History}
At the time when the NSCL built the Coupled Cyclotron Facility (CCF) \cite{CCF}, researchers from MSU and Florida State University (FSU) received a Major Research Instrumentation (MRI) grant from the National Science Foundation (NSF) \cite{SweeperNSF} to construct a large dipole magnet to be used with two existing neutron walls \cite{NeutronWalls} to perform neutron-fragment coincidence experiments. This magnet swept (thus ``Sweeper magnet'') charged fragments to about 40$^\circ$ from the beam axis to allow background-free detection of neutrons at 0$^\circ$. 

\begin{figure}[tb]
 \centering
 \includegraphics[width=6cm]{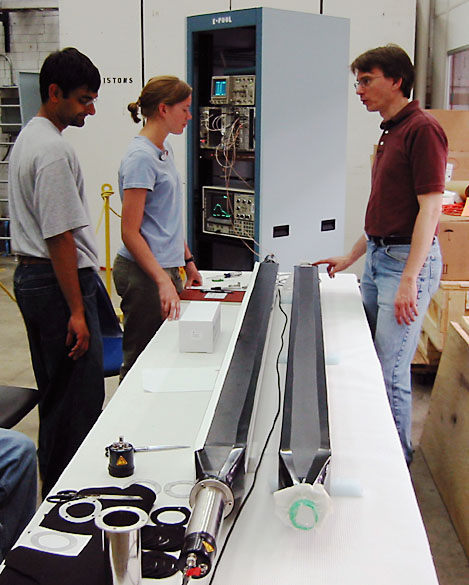}
\caption{Undergraduate students are being trained in the assembly and testing of MoNA detector modules.}
\label{fig:LutherAndStudents}
\end{figure}

Prior to the construction of the ``Sweeper magnet'', the neutron walls were built for lower beam energies and had only a neutron detection efficiency of about 12\% for the energies expected from the CCF. During the 2000 NSCL users meeting, a working group realized the opportunity to significantly enhance the neutron detection efficiency with an array of more layers using plastic scintillator detectors.

Among others, two users from primarily undergraduate institutions (PUIs) attended the meeting: Jim Brown from Millikin University and Bryan Luther from Concordia College who at the time was on sabbatical at the NSCL. They suggested that the modular nature and straight forward construction of such an array would offer great opportunities to involve undergraduate students. 

The idea evolved into a design of an array of nine layers with 16 plastic scintillation detectors each of dimensions 10cm x 10cm x 200cm. In the spring of 2001 a collaboration of MSU, FSU, and eight undergraduate institutions was formed that submitted nine (one for each layer) NSF MRI proposals \cite{MoNAMSUNSF}. The faculty and institutions that submitted the grants were Thomas Baumann, Gregers Hansen, Kirby Kemper, Sam Tabor, and Michael Thoennessen (MSU/FSU), Jim Brown (Millikin University), Paul DeYoung and Graham Peaslee (Hope College), Joe Finck (Central Michigan University), Jerry Hinnefeld (Indiana University South Bend), Ruth Howes (Ball State University), Bryan Luther (Concordia College), Paul Pancella (Western Michigan University\footnote{The Physics Department has a graduate program, but only undergraduate students were involved in the MoNA Collaboration}), and Warren Rogers (Westmont College). The proposals were successful and the NSF awarded funding in the fall of 2001. 

\begin{figure}
 \centering
\includegraphics[width=\linewidth]{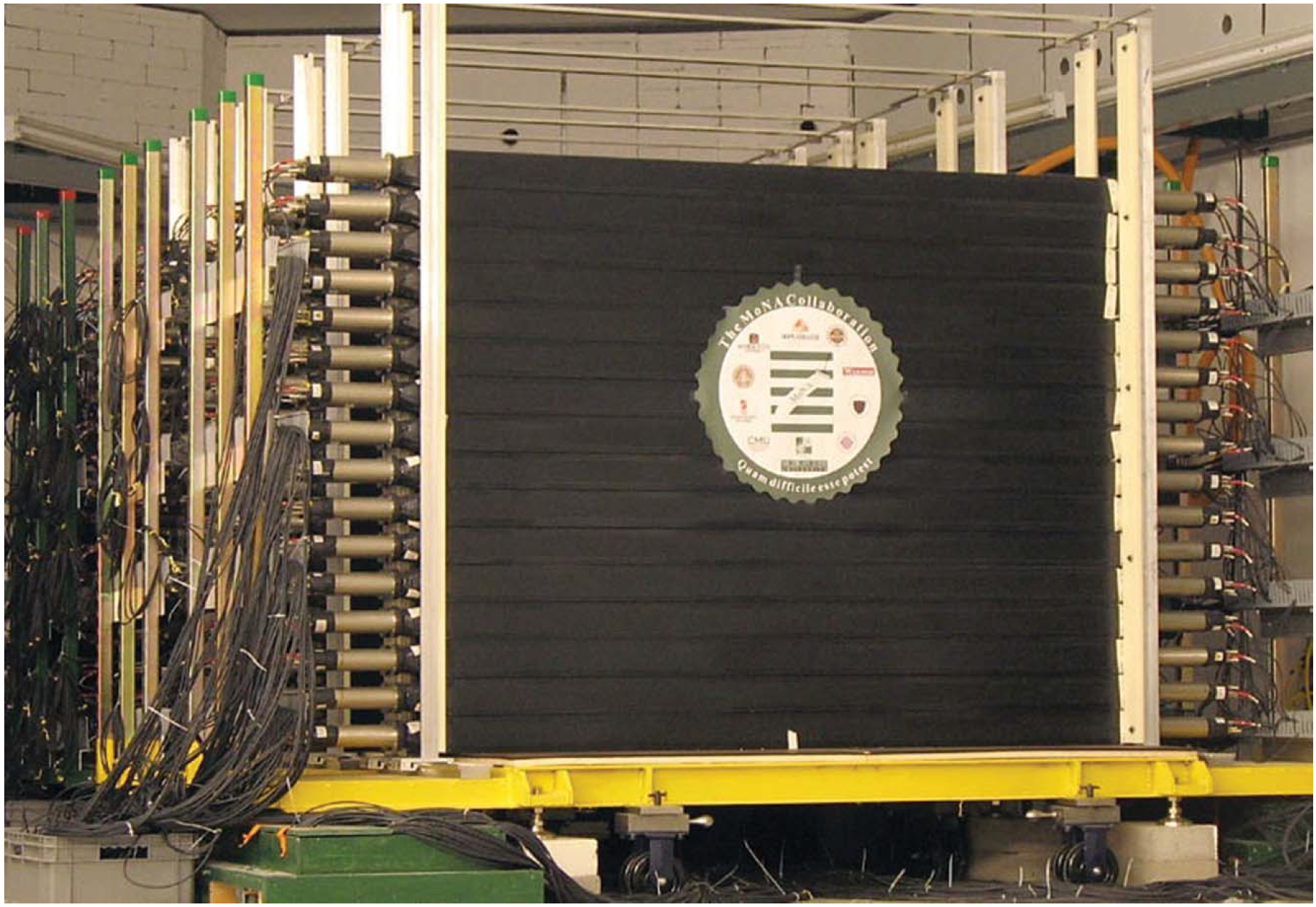}
\caption{The assembled Modular Neutron Array (MoNA) \cite{BrownNPN}.}
\label{fig:MoNA}
\end{figure}

\begin{figure}
 \centering
 \includegraphics[width=\linewidth]{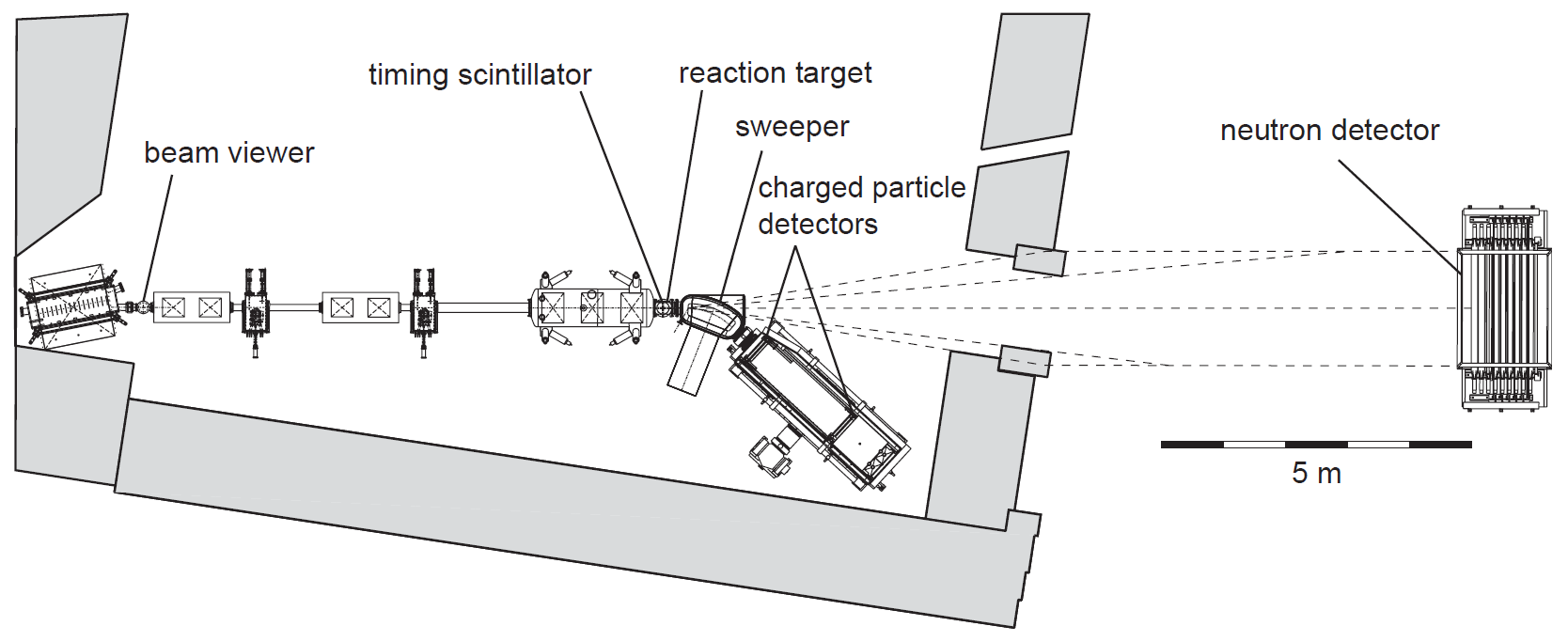}
\caption{MoNA setup in the N4 vault \cite{Bau05}.}.
\label{fig:MoNA-N4}
\end{figure}

Following a detailed design and construction, the first detector array modules were delivered in the summer of 2002. During the following year, all modules were assembled and tested by undergraduate students at their respective schools \cite{AJP} (see figure \ref{fig:LutherAndStudents}). The modules were then transported to the NSCL where they were assembled to form the complete array, the Modular Neutron Array (MoNA) as shown in figure \ref{fig:MoNA}. MoNA was positioned behind an opening in the N4 vault where the target and the Sweeper magnet were located as shown in figure \ref{fig:MoNA-N4}. 

Once the initial construction and commissioning of MoNA was complete, the collaboration began a research program focusing on light neutron-rich nuclides. This provided opportunities for many undergraduate students from the collaborating schools to participate in cutting-edge nuclear physics experiments at one of the world’s leading rare-isotope facilities. Research at the undergraduate institutions was partially funded by the NSF through several RUI grants (Research at Undergraduate Institutions). Undergraduate students were involved in the preparation and running of the experiments as well as in data analysis, simulations, and dissemination of the physics results \cite{MongolHorde}. This approach of involving undergraduates in all aspects of research has continued through today.

\begin{figure}
 \centering
 \includegraphics[width=\linewidth]{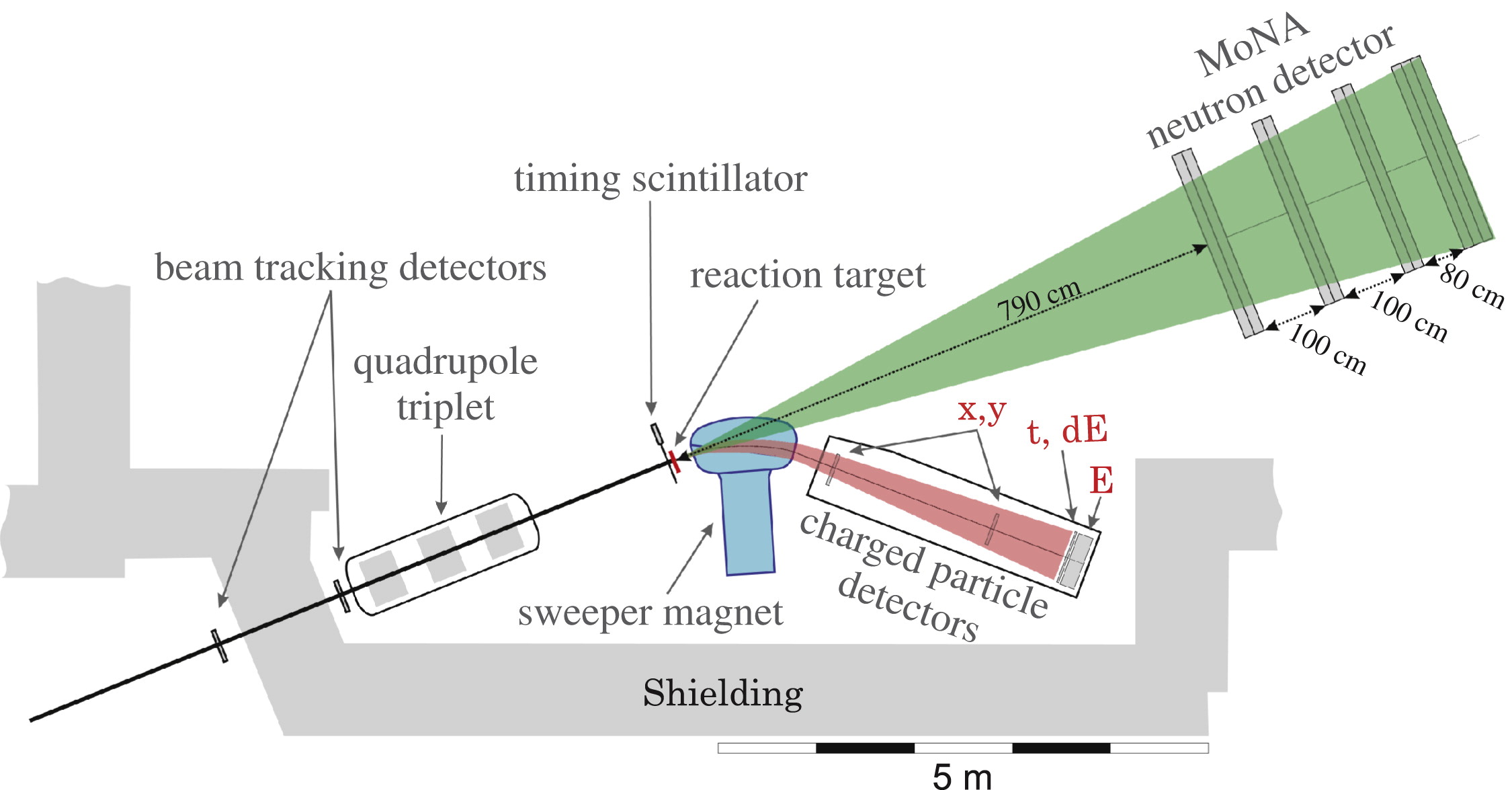}
\caption{MoNA setup in the N2 vault \cite{KOHLEY201259}.}.
\label{fig:MoNA-N2}
\end{figure}

In 2007, the MoNA-Sweeper setup was moved to the newly expanded N2 vault as shown in figure \ref{fig:MoNA-N2}. Again, undergraduate students were essential during the disassembly and reassembly of the array.

\begin{figure}[htb]
 \centering
 \includegraphics[width=0.8\linewidth]{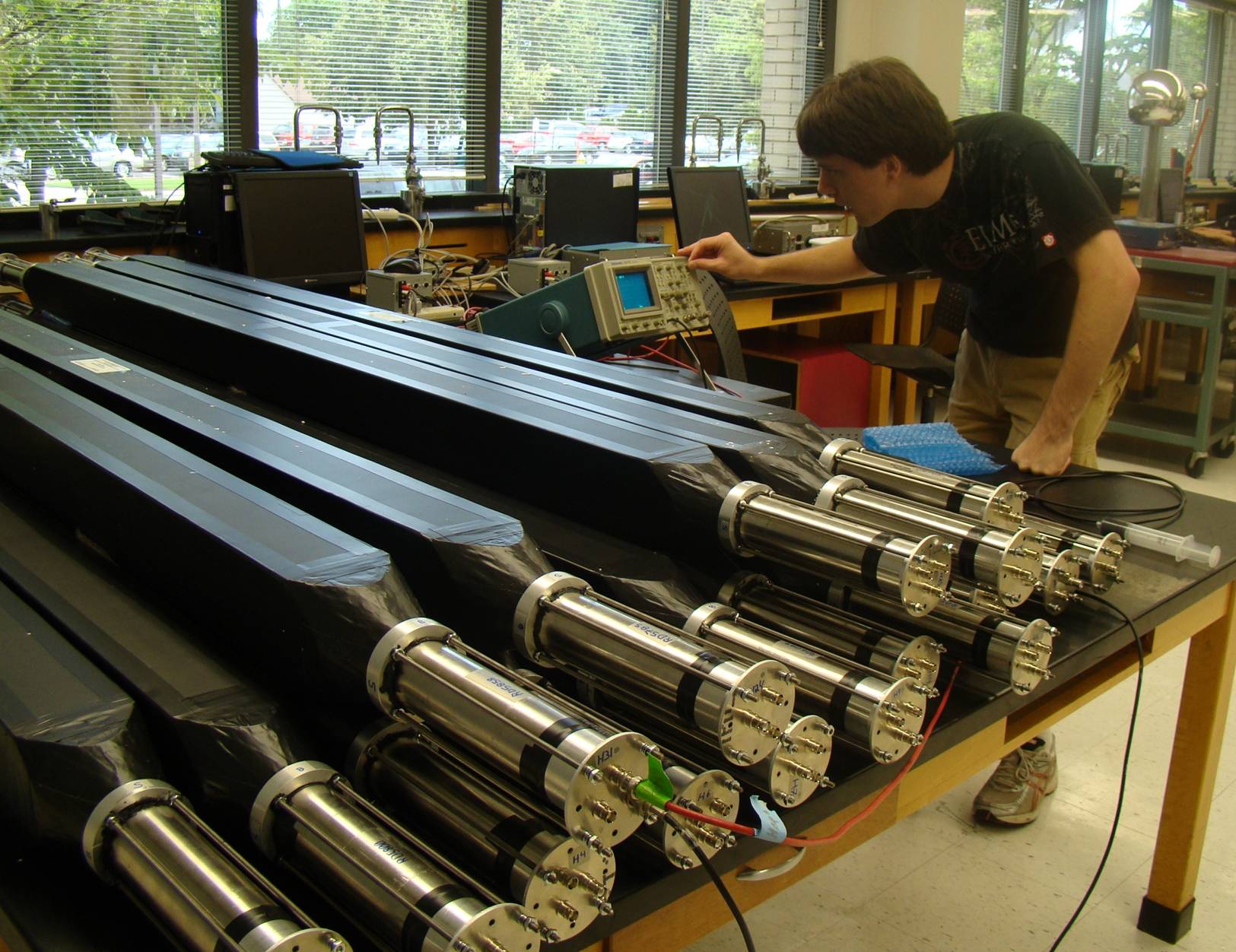}
\caption{Assembled LISA modules being tested.}.
\label{fig:lisafull}
\end{figure}

After performing experiments for eight years, additional detection modules were identified as a way to improve the angular detection acceptance for single and multiple neutron emission. Similar to the original MoNA proposals, the construction of nine additional layers of 16 modules by each institution was proposed for the NSF/MRI program in 2009. It provided an opportunity to expand the collaboration to include additional PUIs. In addition to six proposals from original MoNA members (Hope College, Concordia College, Indiana University South Bend, Westmont College, Central Michigan University, and Wabash College,\footnote{Jim Brown had moved in 2003 from Millikin University to Wabash College.}) three layers were proposed by Sharon Stephenson from Gettysburg College, Deseree Meyer from Rhodes College, and Bob Kaye from Ohio Wesleyan University \cite{LISANSF}, one layer for each new instituion.  The new array was named the Large-area multi-Institutional Scintillator Array (LISA) and it was constructed just like MoNA by undergraduate students during the summer of 2010. 

\begin{figure}[tb]
\centering
\includegraphics[width=\linewidth]{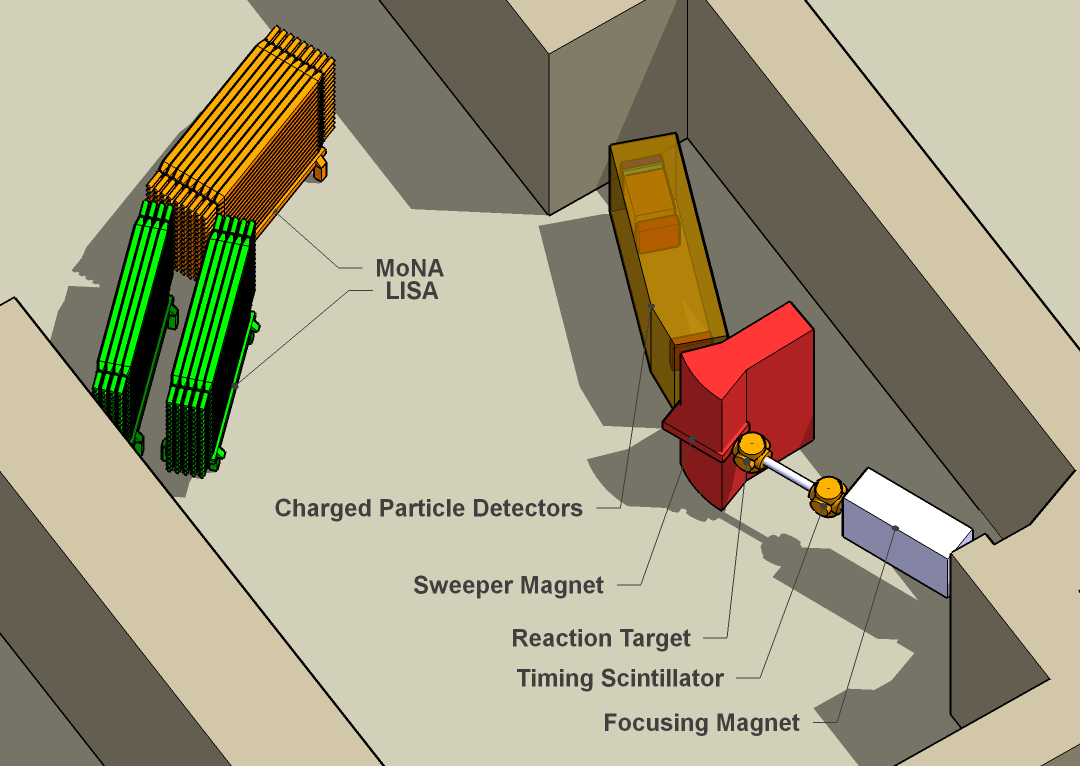}
\caption{Layout of MoNA-LISA in the N2 vault \cite{ActaPP}.}
\label{fig:Vault}
\end{figure}

In addition to testing the modules, undergraduate students carried out various projects at their home institutions such as measurements of the muon-lifetime, cosmic-ray shower size, and $\gamma-\gamma$ correlations (see figure \ref{fig:lisafull}). The modules were installed at the NSCL in January 2011 and the final integration with MoNA and the Sweeper magnet was completed in the summer of 2011 (see figure \ref{fig:Vault}).

During the LISA project, the collaboration saw additional changes to its membership. In 2008, Ruth Howes, who had become the Chair of the Marquette University Physics Department in 2003, retired and became the first emeritus member of the collaboration. Artemis Spyrou (MSU) and Nathan Frank (Augustana College) joined MoNA in 2009.

\begin{figure*}[tb]
 \centering
 \includegraphics[width=0.9\textwidth]{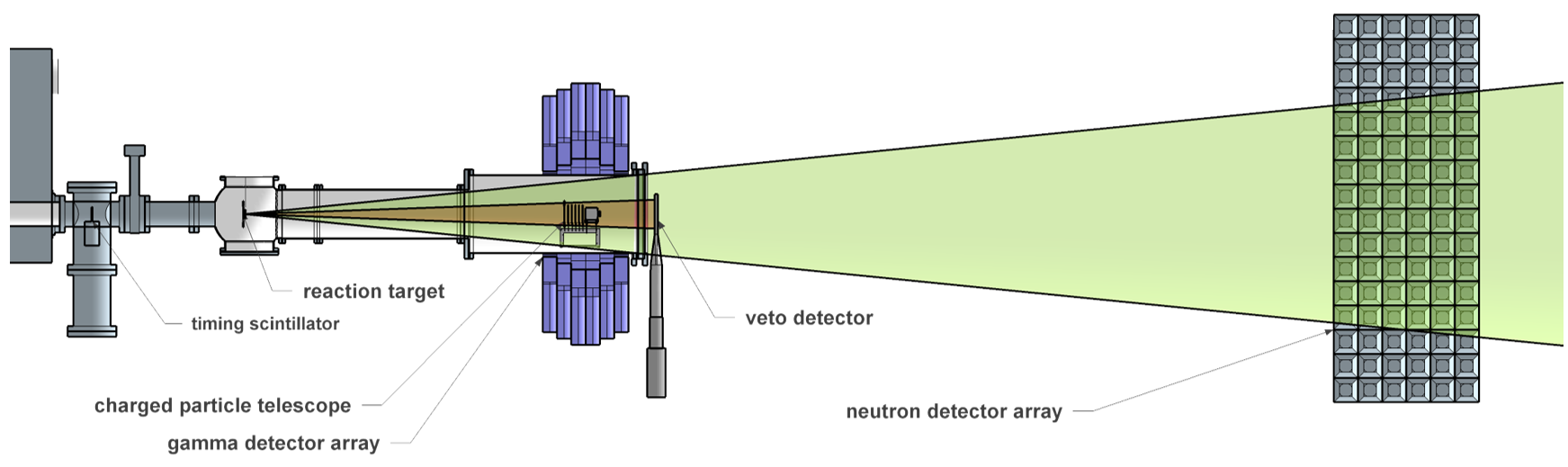}
\caption{Experimental setup in the S2 Vault.}
\label{fig:S2}
\end{figure*}

In 2012, Nathan Frank and Michael Thoennessen together with Paul Gueye from Hampton University received a grant from the Nuclear Science and Security Consortium (NSSC) \cite{NSSC} to construct a segmented target to determine the location of nuclear reactions within the reaction target and thus provide better resolution in decay energy measurements. As a principal investigator of the grant, Paul Gueye officially joined the MoNA Collaboration in 2013. 

Also in 2012, the thick plastic detector to measure the full energy of the charged particles behind the Sweeper was replaced by an array of 16 CsI(Na) scintillation detectors.

Further changes to the membership demonstrate the dynamic evolution of the collaboration. By 2015, the members of FSU, Western Michigan University, and Rhodes College had become inactive in the collaboration. In 2015, Bryan Luther moved to St. Johns College (he returned to Concordia College as Emeritus Professor in 2019) and Warren Rogers became the Blanchard Chair in Physics at Indiana Wesleyan University in 2016. 

In the following years, two members had to leave the collaboration. Michael Thoennessen was appointed the Editor in Chief of the American Physical Society (2017) and Sharon Stephenson accepted a position as a program manager in the Office of Nuclear Physics at the Department of Energy (2019). In the meantime, Anthony Kuchera from Davidson College joined the collaboration, Paul Gueye moved full-time to MSU, and Joe Finck became Emeritus Professor (2018). In 2020, Bob Kaye moved to Westmont College and became inactive. Thomas Redpath (Virginia State University) and Calem Hoffmann (Argonne National Laboratory) joined in 2021 and Belen Monteagudo Godoy (Hope College) and Adriana Banu (James Madison University) in 2022.

At the beginning of 2023, Michael Thoennessen returned to MSU as Emeritus Professor and later in the year Aldric Revel joined the MSU faculty and the MoNA Collaboration as its newest member. 

During this time, the collaboration continued to improve and upgrade its equipment and capabilities. The NSCL CAESAR \cite{CAESAR} array was added to a MoNA/LISA experiment for the first time in 2010 to measure neutrons in coincidence with $\gamma$-rays (in addition to the charged fragments).

In 2015, the MoNA Collaboration sent 16 scintillator bars to the Los Alamos Neutron Science Center (LANSCE) to benchmark the response of plastic scintillators to neutron scattering.
 
Nathan Frank, Paul DeYoung, Jim Brown, and Thomas Baumann received another NSF/MRI award in 2018 to develop a Charged Particle Detector Telescope (CPDT) \cite{CPTelescopeNSF}. The new Si/CsI-based telescope device was designed to increase the efficiency for detection of charged fragments, especially for triple-coincidence measurements of fragments, neutrons, and $\gamma$-rays.

The CPDT was used in the last NSCL experiment before the transition to FRIB, which also required another move (see figure \ref{fig:S2}). Unfortunately, no undergraduate students were involved in the experiment as it was performed during the COVID pandemic. However, undergraduate students were involved in the CPDT development and commissioning as well as in the data analysis.

\begin{figure*}[tb]
 \centering
 \includegraphics[width=0.99\textwidth]{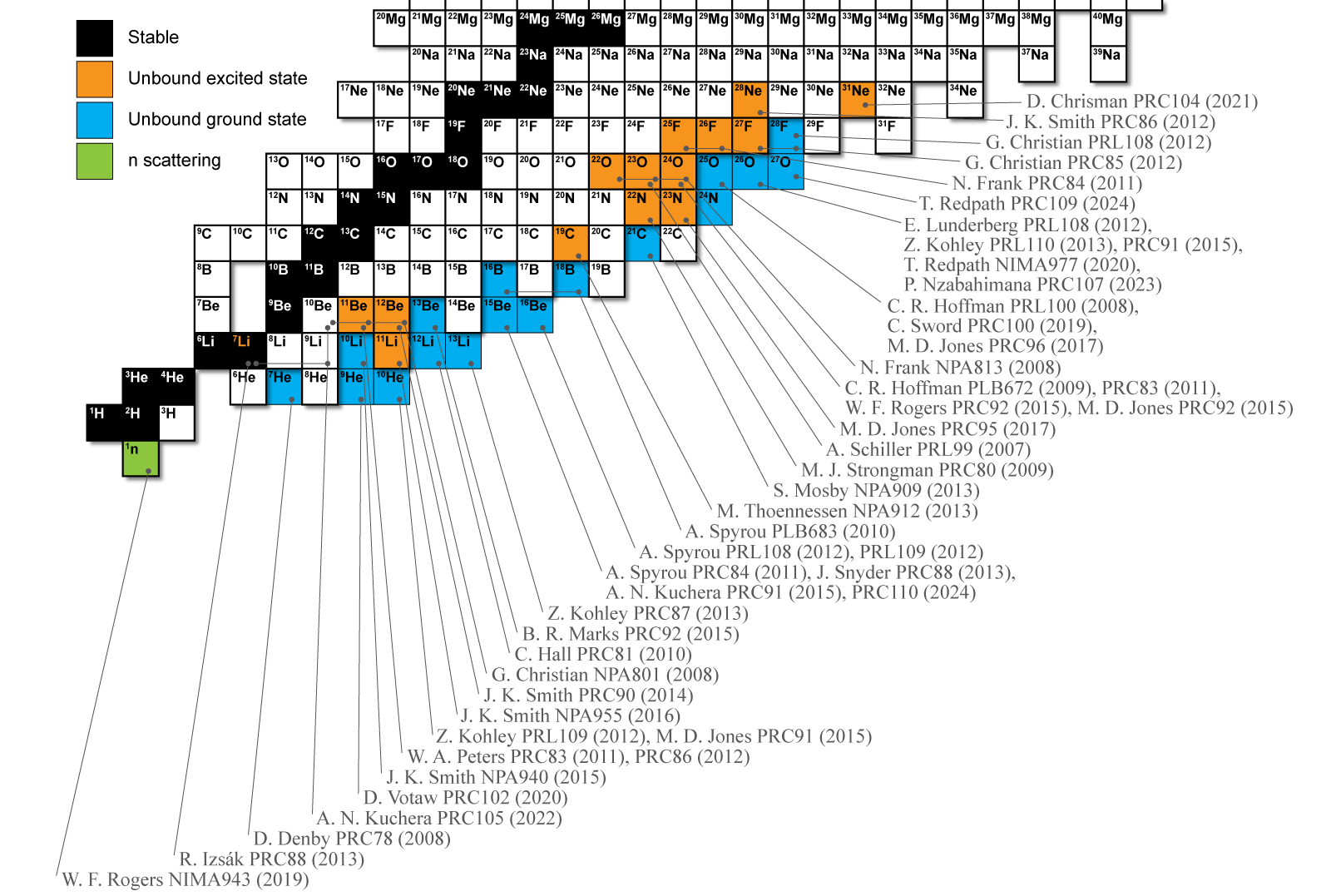}
\caption{Chart of the nuclides for Z $\leq 12$ highlighting nuclides with MoNA publications.}
\label{fig:chartofexperiments}
\end{figure*}

The next major step towards the future was the proposal for the Next Generation neutron (NGn) detector array. The innovative design based on Silicon Photomultipliers (SiPM) which will significantly improve the position resolution and thus the neutron kinetic energy resolution was submitted to the NSF as eight collaborative MRI proposals in the spirit of MoNA and LISA. The NSF approved the proposals from Augustana College, Davidson College, Hope College, Indiana Wesleyan University, James Madison University, MSU, Virginia State University, and Wabash College at the end of 2023 \cite{NGnNSF}. A first test of small scintillator bars with silicon photomultipliers (SiPM) was already performed at the Triangle Universities Nuclear Laboratory (TUNL).

The collaboration has seen much change over the years but has maintained its mission of performing cutting-edge research and mentoring the next generation of physicists. The NGn project will be completed by 2026 for exciting new experiments at FRIB.

\section{Physics with MoNA}

The experiments of the MoNA Collaboration focused on the exploration of nuclides at and beyond the neutron dripline. Figure \ref{fig:chartofexperiments} shows a section of the chart of nuclides indicating nuclides that were studied during the first 25 years of MoNA. A general overview of nuclear structure experiments along the neutron drip line can be found in Ref. \cite{Baumann_2012}. Most experiments extracted resonance parameters of ground or excited states utilizing the invariant mass method.

\subsection{Invariant mass method}
Invariant mass measurements require knowledge of the kinetic properties of all decay products, \textit{i.~e.} the neutron (or neutrons, indexed n) and the charged fragment (indexed f). For a one-neutron unbound system the invariant mass ($M$) can be calculated from the energies ($E_\text{n}$ and $E_\text{f}$) and momentum vectors ($\vec{p}_\text{n}$ and $\vec{p}_\text{f}$): 

\begin{displaymath}
M=\sqrt{m^2_\text{f}+m^2_\text{n}+2(E_\text{f} E_\text{n}/c^4-\\\vec{p}_\text{f} \cdot \vec{p}_\text{n}/c^2)}
\end{displaymath}

The decay energy $E_\text{d}$ corresponds to the difference between the invariant mass of the unbound system and the sum of the masses of the decay products (utilizing $E = mc^2)$:

\begin{displaymath}
E_\text{d}=Mc^2
-(m_\text{f} + m_\text{n})c^2
\end{displaymath}

The energy and momentum of the neutrons are deduced from the time-of-flight from the target to the interaction point in a MoNA bar, which is determined from the bar location and the time difference between the two ends of the bar \cite{Bau05}. The trajectories of the fragments are measured after the Sweeper magnet by two Cathode Readout Drift Chambers, and the energy and momentum are reconstructed using the magnetic field map of the Sweeper \cite{FRANK20071478}. Additional detectors after the Sweeper magnet aid in charged particle identification.

\begin{figure}[tb]
\centering
\includegraphics[width=\linewidth]{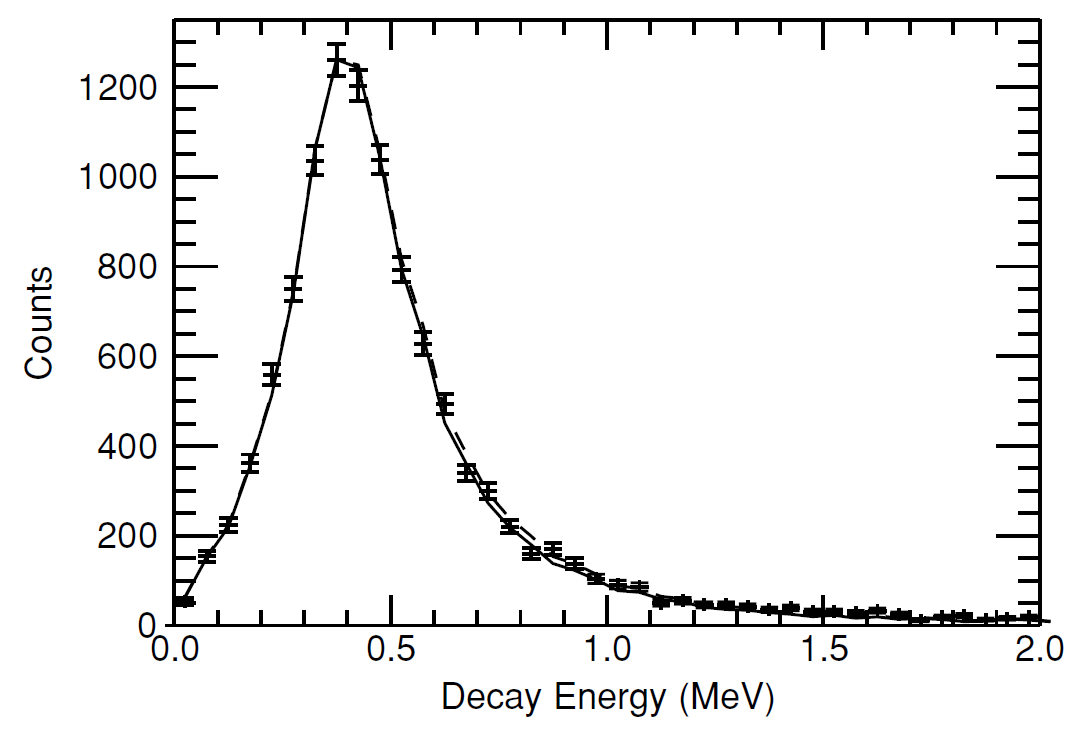}
\caption{The reconstructed decay-energy spectrum for the neutron-unbound ground state in $^7$He \cite{PhysRevC.78.044303}.}
\label{fig:ReconstructedDecayEnergy}
\end{figure}

Figure \ref{fig:ReconstructedDecayEnergy}
shows the reconstructed decay energy of $^{7}$He from one of the first MoNA experiments \cite{PhysRevC.78.044303}.

\subsection{Discovery of neutron-unbound \\isotopes}
\label{Discovery}
Expanding the chart of the nuclides to the limit of stability has been a long-term goal in nuclear physics. However, the question of how many neutrons can be bound by an element is still only known up to sodium. Although some properties of a few nuclides beyond the dripline have been known for a while, many have only been accessible with the availability of radioactive beams. The MoNA/Sweeper setup at the CCF was an ideal tool to discover and measure the properties of isotopes of light elements beyond the dripline. 

Over the last 25 years, MoNA has discovered seven new unbound isotopes between lithium and fluorine. First, $^{25}$O was discovered in 2008 \cite{PhysRevLett.100.152502}, followed by $^{12}$Li \cite{PhysRevC.81.021302} and $^{18}$B \cite{SPYROU2010129} in 2010, $^{16}$Be \cite{PhysRevLett.108.102501,
PhysRevLett.109.239202}, $^{26}$O \cite{PhysRevLett.108.142503}, and $^{28}$F \cite{PhysRevC.85.034327,PhysRevLett.108.032501} in 2012, and $^{15}$Be \cite{PhysRevC.88.031303} in 2013. The ground state of $^{13}$Li was measured for the first time also in 2013 \cite{PhysRevC.92.054320}.

\begin{figure}[tb]
\centering
\includegraphics[width=\linewidth]{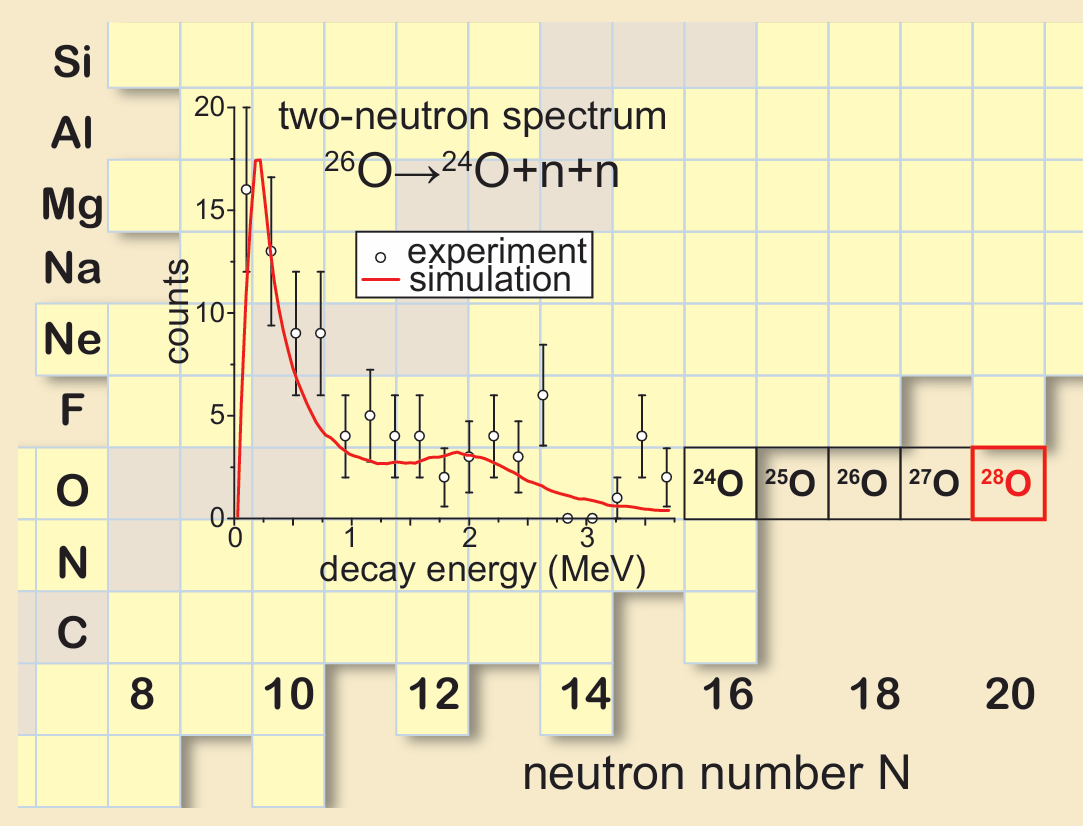}
\caption{Graphics from the 2015 DOE/NSF Long Range Plan highlighting the discovery of $^{26}$O \cite{LRP2015}.}
\label{fig:LRP26O}
\end{figure}

The discovery of $^{26}$O was critical as essentially all theoretical models had predicted it to be bound with respect to one- and two-neutron emission. The observation that $^{26}$O was unbound confirmed that $^{24}$O was the last bound oxygen isotope, which posed a difficult question for the theories: How does one additional proton bind six additional neutrons (the heaviest bound fluorine isotope is $^{31}$F)? These data were highlighted in the 2015 DOE/NSF Long Range Plan \cite{LRP2015} (see figure \ref{fig:LRP26O}).

The observation of $^{16}$Be corresponded to the first observation of a dineutron emission from the ground state of an unbound nucleus \cite{PhysRevLett.108.102501,PhysRevLett.109.239202}.

$^{15}$Be was discovered in a (d,p) reaction after a search with a two-proton removal reaction from $^{17}$C was unsuccessful \cite{PhysRevC.84.044309}. It was observed that the two-proton removal reaction populated a state in $^{15}$Be, which was unbound with respect to three-neutrons decaying to $^{12}$B \cite{PhysRevC.105.034314}. Such a decay branch was not observed in the (d,p) reaction \cite{PhysRevC.91.017304}, which demonstrates reaction selectivity and the importance of populating unbound nuclides using different reactions.

The search for $^{21}$C in the one-proton removal reaction from $^{22}$N yielded no evidence of $^{21}$C and set a limit for the scattering length of a possible s-wave of $|a_s| <$ 2.8~fm \cite{MOSBY201369}. Similarly, in a search for $^{12}$He only a lower limit of about 1 MeV was determined for a possible resonance in $^{12}$He \cite{PhysRevC.91.044312}.

\subsection{Nuclear structure at the dripline\\}
Along the neutron dripline where the neutron binding energy becomes zero, the relatively small enhancement of the total binding energy of paired neutrons has an important effect. The stability of nuclei with even numbers of neutrons $N$ compared to their neighbors with odd numbers creates a saw-tooth pattern in which the heaviest odd-$N$ isotopes of a given element are neutron-unbound while heavier isotopes with an even number of neutrons can be bound. A well-known example is $^{10}$Li (unbound) and $^{11}$Li (bound). The properties of these alternating neutron-unbound nuclides provide important insights into the neutron-nucleus interaction far from stability. 

The structure of nuclides far from stability can be significantly different from stable nuclides. Two prominent examples studied with MoNA are the level inversion of the $s$ and $p$ orbits for neutron-rich $N = 8 $ isotones, and nuclear structure effects occurring in the transition across the $Z = 8$ shell.

\subsubsection*{Disappearance of the $\mathbf{N = 8}$ shell}
The inversion of the $s$ and $p$ levels had already been observed in $^{11}$Be in 1960 \cite{PhysRevLett.4.469} and was the first evidence for the disappearance of the $N = 8$ shell in neutron-rich isotones. Since then, $N = 7$ and $8$ beryllium, lithium, and helium isotopes have been studied extensively to understand the underlying nuclear forces responsible for the inversion.

MoNA has contributed to these studies by measuring the properties of excited unbound states in $^{12}$Be \cite{PhysRevC.90.024309}, $^{11}$Be \cite{PhysRevC.83.057304,PhysRevC.86.019802}, $^{11}$Li \cite{SMITH201627} and $^{10}$Li \cite{SMITH2015235}.

The experiments again demonstrated the selectivity of different reactions. Reactions utilizing halo beams (for example $^{11}$Be or $^{11}$Li) preferentially populated $s$ wave configurations which were not present with non-halo beams \cite{PhysRevC.102.014325}.

The influence of halo beams on the decay energy resonance had also been attributed to a discrepancy in the ground state resonance of $^{10}$He \cite{PhysRevC.77.034611,PhysRevC.90.024610}. However, that explanation was not substantiated by a detailed analysis of the population of $^{10}$He from the $-2p2n$ reaction from $^{14}$Be \cite{PhysRevLett.109.232501,
PhysRevC.91.044312}.

\subsubsection*{Transition across the $\mathbf{Z = 8}$ shell}
Several interesting effects occur at the transition across the $Z = 8$ shell for neutron-rich nuclides. New subshells emerge at $N = 14$ and $N = 16$ and another level crossing generates an island of inversion centered around $^{31}$Na. In this region, neutrons fill orbitals in the $pf$-shell before completing the $sd$-shell. This also leads to the disappearance of the $N = 20$ shell. Finally, the puzzle of why adding one proton outside the $Z = 8$ shell binds six additional neutrons has already been mentioned (see Section \ref{Discovery}).

Again, MoNA has been instrumental in exploring these effects and the understanding of the underlying physics. The observation of the unbound 5/2$^+$ state in $^{23}$O established the $N = 14$ subshell \cite{PhysRevLett.99.112501,FRANK2008199}, which was subsequently also observed in $^{22}$N \cite{PhysRevC.80.021302} as well as in $^{19}$C \cite{THOENNESSEN20131}.

\begin{figure}[tb]
\centering
\includegraphics[width=\linewidth]{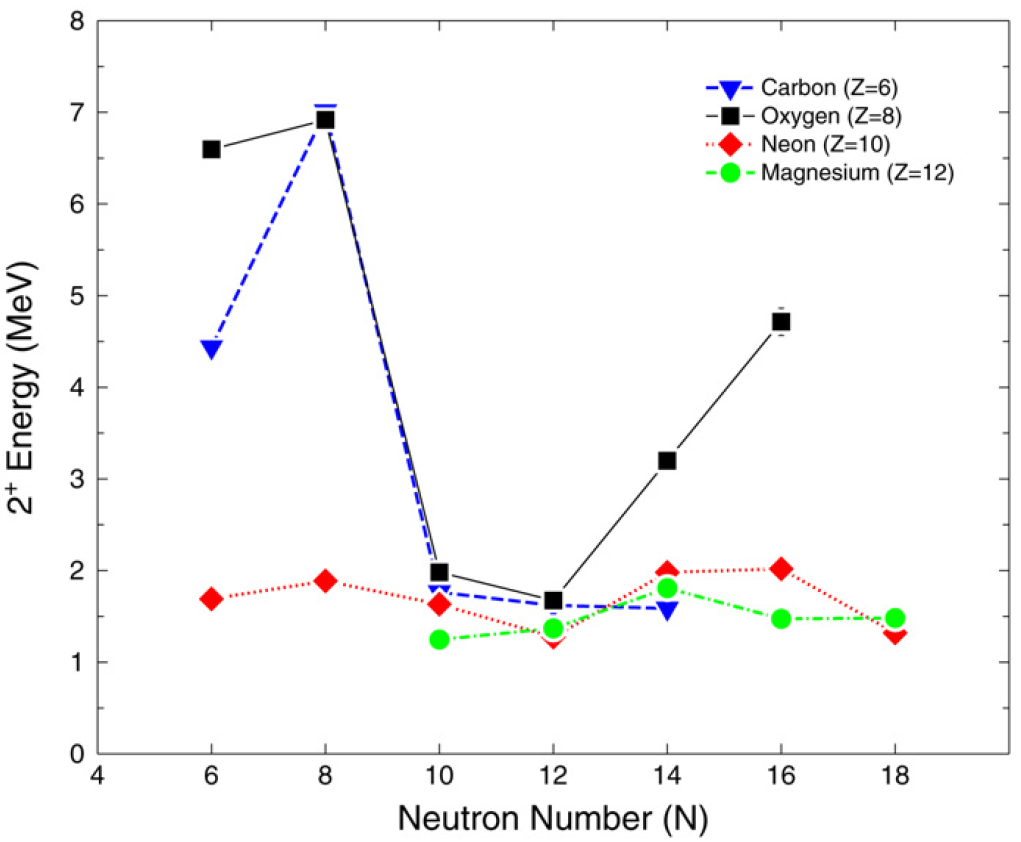}
\caption{Energies of the first 2$^+$ state in even $Z$ = 6 to 12 isotopes as a function of neutron number. The MoNA measurement of $^{24}$O demonstrated the large shell gap at $N = 16$ \cite{HOFFMAN200917}.}
\label{fig:ShellGap}
\end{figure}

The size of the $N = 16$ shell gap was determined from the decay energy of unbound $^{25}$O \cite{PhysRevLett.100.152502,PhysRevC.96.054322} and the excitation energy of the first excited state in $^{24}$O \cite{HOFFMAN200917,PhysRevC.92.034316} (see figure \ref{fig:ShellGap}). The shell gap decreases in lighter isotones as deduced from the excitation energy of the first excited state in $^{23}$N \cite{PhysRevC.95.044323}.

The first measurement of the decay energy of the unbound ground state of $^{28}$F established the lower boundary of the island of inversion \cite{PhysRevLett.108.032501,PhysRevC.85.034327}. This boundary was confirmed by the non-observation of $^{27}$O produced by a two-proton removal reaction from $^{29}$Ne \cite{PhysRevC.109.054325}. Additional measurements of unbound levels in other fluorine \cite{PhysRevC.84.037302} and neon \cite{PhysRevC.86.057302,PhysRevC.104.034313} isotopes contributed to the further theoretical understanding of the island of inversion.

The first observation of the two-neutron decay of $^{26}$O \cite{PhysRevLett.108.142503} confirmed previous unsuccessful searches for the existence of bound $^{26}$O \cite{PhysRevC.41.937,PhysRevC.53.647,PhysRevC.72.037601}. Further data analysis indicated that due to the extremely low decay energy, the ground state might be very long-lived, possibly representing the first occurrence of two-neutron radioactivity \cite{PhysRevLett.110.152501}. Two novel techniques to search for neutron radioactivity with halflives between 1~ps and 1~ns were proposed \cite{THOENNESSEN2013207}. 

\subsection{Identification of multi-neutron\\ emission}

The first demonstration that MoNA could identify the decay of two neutrons was the decay of a highly excited state in $^{24}$O \cite{PhysRevC.83.031303}. From the energy correlation of the two-body subsystems ($E_{fn_1}$ and $E_{fn_2})$ it was determined that the decay proceeded via the first excited state of $^{23}$O. This conclusion was later confirmed by adding the spatial correlations in the Jacobi coordinates \cite{PhysRevC.92.051306}. The Jabobi analysis was also used in the discovery of the dineutron decay of $^{16}$Be \cite{PhysRevLett.108.102501} and the two-neutron decay of $^{26}$O \cite{PhysRevC.91.034323}.

In the first attempt to push the multi-neutron capabilities even further, the search for the three-neutron decay of $^{15}$Be in a two-proton removal reaction from $^{17}$C did not yield any events \cite{PhysRevC.91.017304}. However, a few years later this decay was observed in the $^{14}$Be(d,p)
transfer reaction \cite{PhysRevC.110.064302}, which again demonstrated the importance of utilizing different reaction mechanisms. In the meantime, the sequential three-neutron decay was observed for the first time, identifying a highly excited state in $^{25}$O \cite{PhysRevC.100.034323}. 

\subsection{Reaction mechanisms}
Most of the MoNA experiments used single- or multiple-proton and neutron removal reactions. As mentioned above, populating the states with different reaction mechanisms yields additional nuclear structure information. Other reactions explored by MoNA were for example inelastic scattering \cite{PhysRevC.92.051306}, charge exchange \cite{PhysRevC.92.054320,PhysRevC.84.037302}, neutron pick-up or (d,p) \cite{PhysRevC.88.031303,PhysRevC.110.064302} and fragmentation \cite{CHRISTIAN2008101} reactions.

In addition to exploring nuclear structure effects at and beyond the neutron dripline, a few MoNA experiments focused on other goals. Coulomb dissociation reactions can be used to measure the cross section for the inverse neutron capture reaction. MoNA explored the reaction $^8$Li($\gamma$,n)$^7$Li to extract the cross section for the capture reaction $^7$Li(n,$\gamma$)$^8$Li \cite{PhysRevC.88.065808}, which is important for understanding nuclear reactions in the Big Bang.

Another example was the use of fragmentation reactions of neutron-rich radioactive beams to constrain the symmetry energy of the nuclear equation of state. Neutron multiplicities were measured in coincidence with charged fragments following the fragmentation of $^{32}$Mg and the data were compared with a constrained molecular dynamics model to demonstrate that the observables were sensitive to the symmetry energy \cite{PhysRevC.88.041601}.

\subsection{Understanding neutron interactions with matter}

\begin{figure}[tb]
\centering
\includegraphics[width=\linewidth]{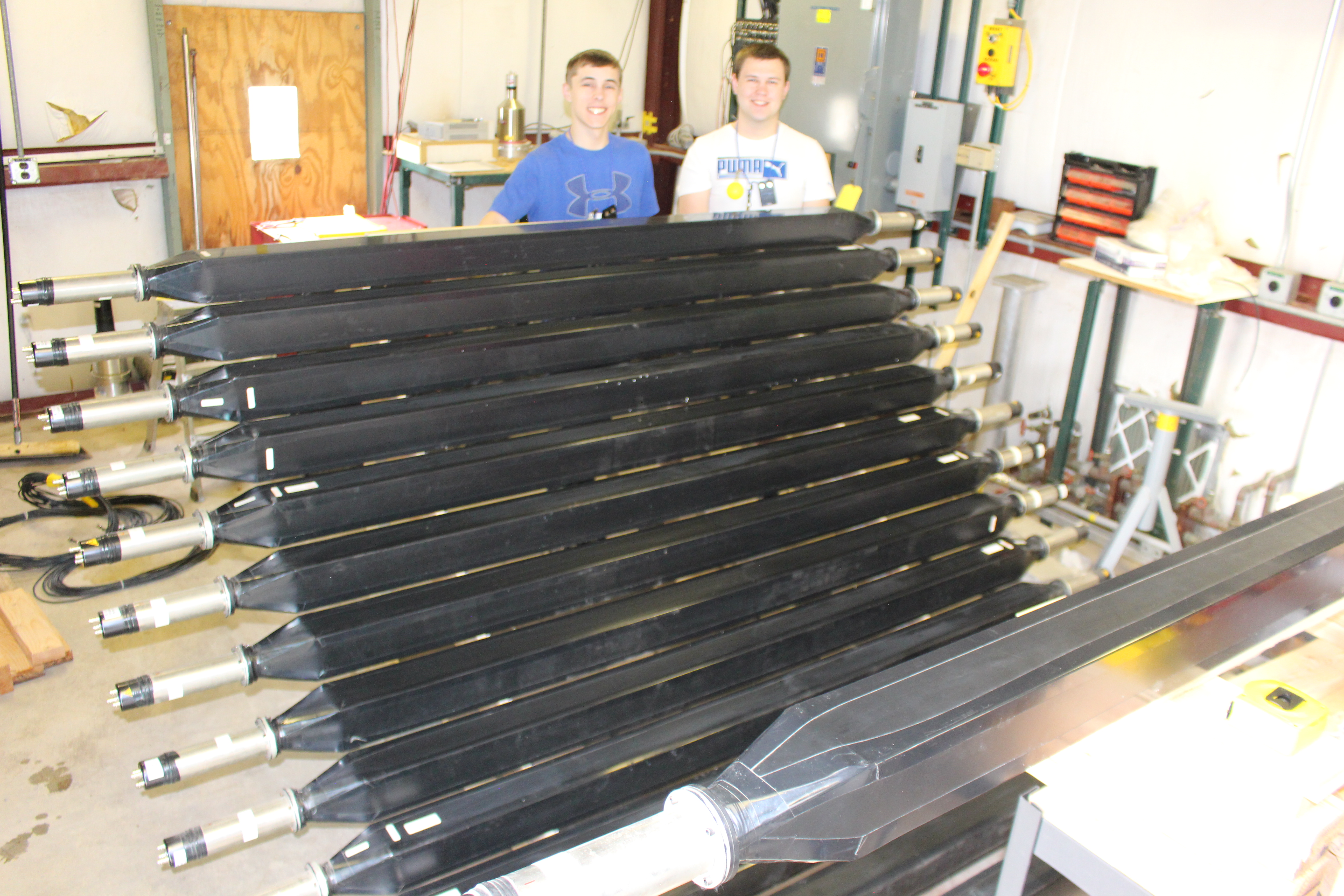}
\caption{MoNA detector setup at the Los Alamos Neutron Science Center (LANSCE).}
\label{fig:LANL}
\end{figure}

Neutrons are detected through their nuclear interactions with matter. A single neutron can interact multiple times with a detector, so it is important to understand these interactions to distinguish real two-neutron interactions from a single neutron interacting twice (or more). From the beginning, MoNA investigated neutron-matter interaction by comparing experiments with simulations. The use of passive converters to enhance the neutron detection efficiency influenced the design of the array \cite{BAUMANN200325}. 

MoNA has relied on the Monte Carlo simulation package GEANT4 to compare data to theory. However, the standard physics processes modeling neutrons did not reproduce experimental observables for intermediate neutron energies ($\sim 20-$300~MeV). MoNA demonstrated that the custom interaction model MENATE$\_$R resulted in excellent agreement with measured neutron multiplicities and deposited energies for neutron energies of $\sim$50 MeV\cite{KOHLEY201259}.

In 2015, the MoNA Collaboration sent sixteen scintillator bars to LANSCE to further benchmark neutron interactions in GEANT4 with the white neutron beam since it produces neutrons of energies up to 800 MeV. Although simulations with MENATE$\_$R performed better, challenges remained to construct physics based filtering algorithms that are required to accurately extract true two- or more neutron events \cite{ROGERS2019162436}. 

Follow-up experiments were performed with different array configurations (see figure \ref{fig:LANL}) and utilizing diamond detectors as active targets. 

Finally, data from two-neutron decay of $^{26}$O were recently analyzed using a deblurring method and a machine learning model based on a deep neural network \cite{PhysRevC.107.064315}.

\section{Experimental devices}

\subsection{Sweeper magnet}

\begin{figure}
 \centering
\includegraphics[width=0.8\linewidth]{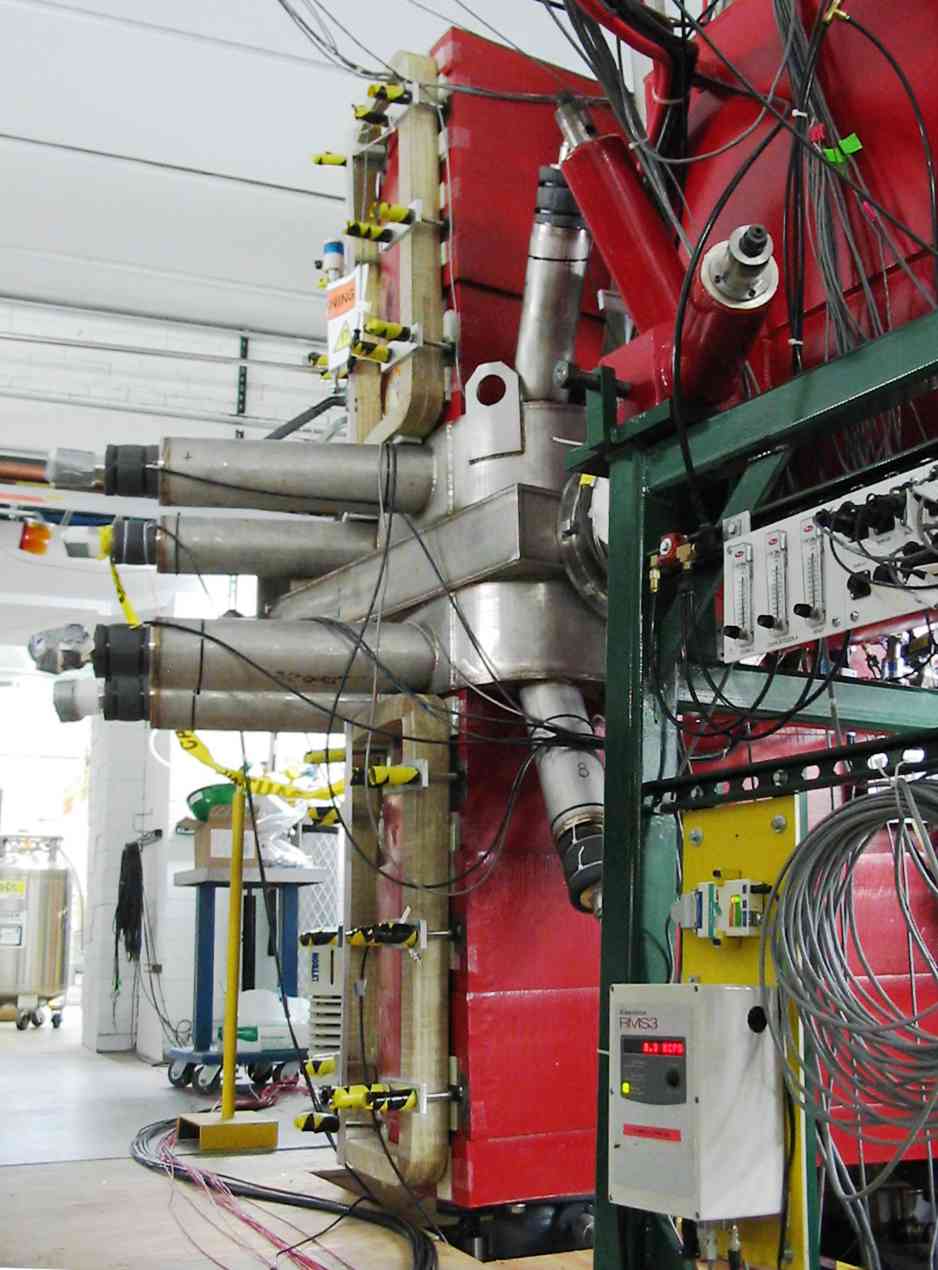}
\caption{The FSU/MSU Sweeper magnet assembled in the N4 vault.}
\label{fig:SweeperMagnet}
\end{figure}

The Sweeper magnet shown in figure \ref{fig:SweeperMagnet} is a large-gap dipole magnet developed and built at the National High Magnetic Field Laboratory at Florida State University \cite{Zel00,Pre01,Tot02,Bir04,Bir05}. The superconducting magnet deflects charged particles by a 43-degree angle up to a rigidity of 4~Tm to separate neutrons, charged reaction products, and the non-reacting beam particles. The vertical gap between the pole tips measures 14~cm and a large neutron window enables the neutrons from the reaction target placed in front of the Sweeper to reach MoNA and LISA, typically placed at or near 0$^{\circ}$ from the incoming beam direction.

A vacuum box directly behind the Sweeper contains two Cathode Readout Drift Chambers followed by an ionization chamber, a thin plastic scintillator, and an array of 16 CsI(Na) crystals. These detectors measure the position and angle, the energy-loss, the time-of-flight, and the total remaining energy, respectively.

\subsection{Modular Neutron Array (MoNA)}

The Modular Neutron Array (MoNA) is a large-area, high-efficiency neutron detector array consisting of 144 10cm x 10cm x 200cm plastic scintillation bars with photomultiplier tubes (PMT) attached to each end. In its standard configuration, MoNA has an active area of 2.0~m wide by 1.6~m tall. It measures both the position and time of neutron events with multiple-hit capability. The position is determined by the bar location and the time difference between the two PMT signals of the bar. The neutron time-of-flight is extracted from the average time of the PMT signals and a start detector located close to the target.
The energy of a neutron is then calculated from the time-of-flight, and its momentum vector is reconstructed from the time-of-flight and the calculated position \cite{Lut03,Bau05}.

The detection efficiency of MoNA is maximized for the high-beam velocities available at the NSCL (now FRIB). For neutrons ranging from 50 to 250~MeV in energy, it is designed to have a neutron detecting efficiency of up to 70\%.

The modular design of the array allows for different configurations to optimize the acceptance and efficiency of the array for the specific reaction of interest.

\subsection{Large-area multi-Institutional\\ Scintillator Array (LISA)}

\begin{figure}[tb]
 \centering
\includegraphics[width=\linewidth]{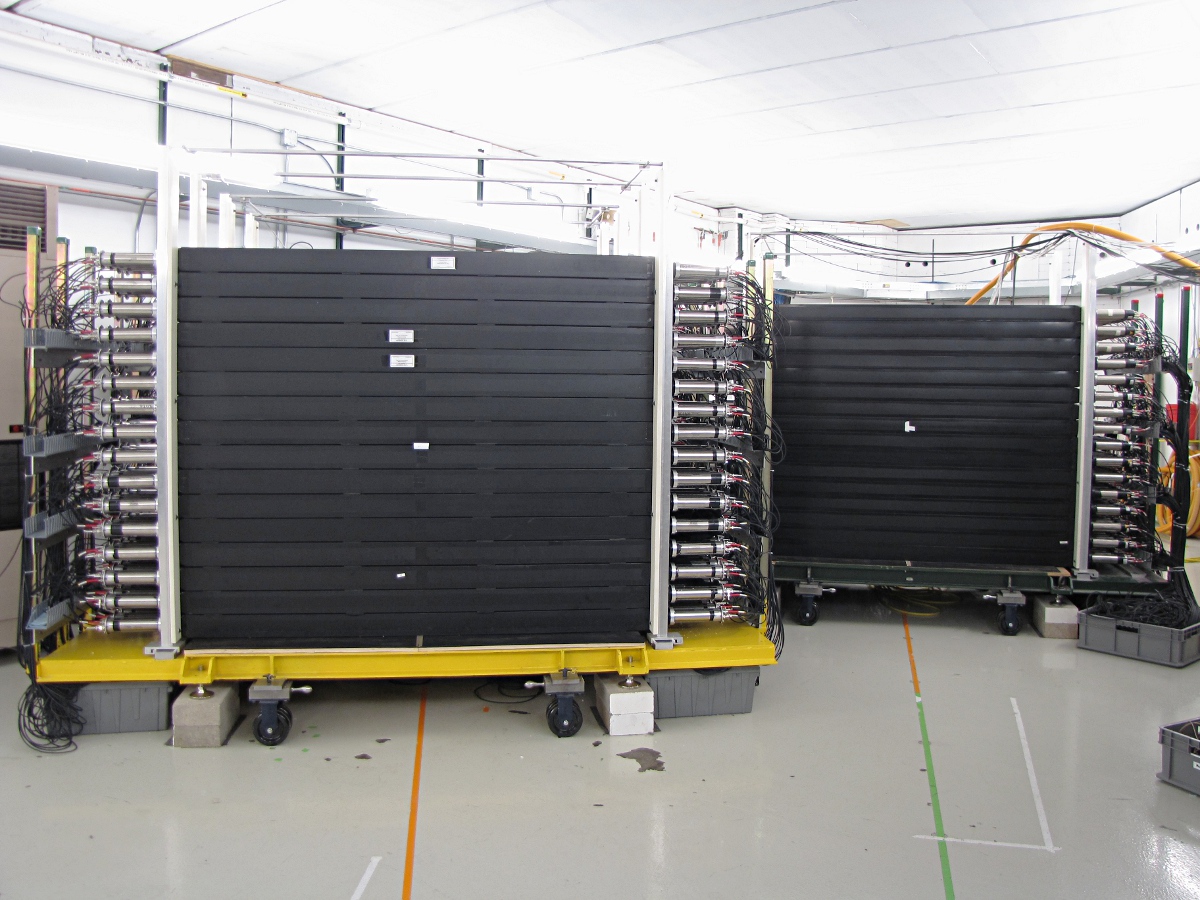}
\caption{The Modular Neutron Array (right) and Large-area multi-Institutional Scintillator Array (left) (MoNA-LISA) in the N2 vault.}
\label{fig:MoNA-LISA}
\end{figure}

LISA is essentially a copy of MoNA. It also consists of 144 modules with the same dimensions. It was constructed to enhance the overall neutron detection capabilities (see figure \ref{fig:MoNA-LISA}). The combined MoNA-LISA array can be configured for additional angle coverage or additional efficiency.

\subsection{Segmented active target}
The low beam intensity available for neutron-rich beams far from stability requires thick reaction targets to collect sufficient statistics to measure unbound ground or excited states of nuclides. However, the depth uncertainty of the nuclear reaction within the target decreased the decay energy resolution.

\begin{figure}[!hb]
\centering
\includegraphics[width=0.4\linewidth]{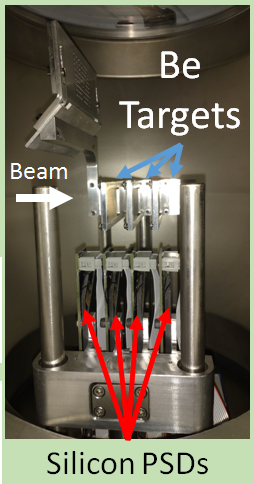}
\caption{The segmented target inside the scattering chamber. The silicon position sensitive detectors (PSD) are in the retracted position.}
\label{fig:seg_targ_setup2}
\end{figure}

To overcome these limitations, a segmented target consisting of alternating layers of silicon detectors (62~mm x 62~mm x 140~$\mu$m) and passive beryllium targets (600~mg/cm$^2$ each) was constructed as shown in figure \ref{fig:seg_targ_setup2}. The energy loss of the secondary beam and charged reaction product nuclei were measured in each detector to determine event-by-event the beryllium target in which the nuclear reaction occurred. This determination provides a means to keep the resolution in decay energy measurements constant while using a thicker beryllium target to increase statistics. In addition, the readout from each corner of the detector provided a position measurement at the target position \cite{REDPATH2020164284}.

\subsection{Auxiliary detectors}

\subsubsection*{CAESium-iodide scintillator ARray \\(CAESAR)}

Decay energies extracted from coincidence experiments between charged fragments and neutrons using the invariant mass method correspond to the excitation energy of the unbound state only if the final fragment is in its ground state. If the final fragment has bound excited states, the measurement is ambiguous. In these cases, additional $\gamma$-ray coincidence measurements are necessary to resolve these ambiguities. 

CAESAR is a highly efficient, segmented scintillation array designed for in-beam$\gamma$-ray spectroscopy experiments utilizing fast beams of rare isotopes \cite{CAESAR}. The crystal shapes of the 192 individual modules allow for a close-packed geometry around the target, yielding high solid angle coverage. The array was used in experiments to study neutron-rich fluorine isotopes \cite{PhysRevLett.108.032501,PhysRevC.85.034327}, measure $\gamma$ rays from nucleon removal reactions of $p$-shell nuclei \cite{PhysRevC.105.034314}, and most recently in the search for the population of the $^{12}$Be second 0$^+$ state in the decay of $^{13}$Be.

\subsubsection*{Liquid hydrogen target}

While most MoNA experiments were performed using thick beryllium targets, the selective population of specific states required different reaction mechanisms and thus different targets. Neutron pickup reactions from deuterium (d,p) can populate states that are inaccessible with proton removal reactions.

Liquid hydrogen targets are ideal for these reactions. They have a higher hydrogen density than polypropylene targets and do not have any background from reactions on carbon.

For the search for unbound excited states in $^{25}$O in the reaction $^{24}$O(d,p) MoNA installed the Ursinus College Liquid Hydrogen Target which was developed at the NSCL in 2010 and filled it with liquid deuterium \cite{PhysRevC.96.054322}.

\subsection {Charged Particle Detector \\Telescope (CPDT)}

A new detection system to enable efficient detection of charged particles, neutrons, and $\gamma$ rays was constructed in 2019.

\begin{figure}[!b]
\centering
\includegraphics[width=\linewidth]{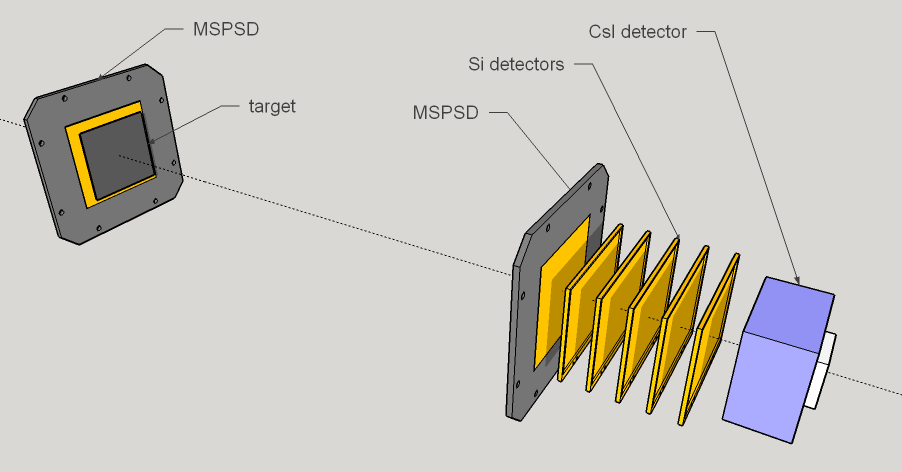}
\caption{Components of the charged particle telescope. (Beam enters from the left.)}
\label{fig:nathandetector}
\end{figure}

The CPDT consists of a stack of silicon detectors backed by a CsI crystal (see figure \ref{fig:nathandetector}). The secondary beam passes through one silicon position-sensitive detector (MSPSD), enters the reaction target, passes through another MSPSD, followed by five energy-loss silicon detectors, and is then deposited in the CsI energy detector that is read out by a silicon pin diode (see figure \ref{fig:nathandetector}). The energy loss measurements and the total energy measurement provide the charged particle identification for the system. 

The CPDT was used for the first time in an experiment to search for the population of the second 0$^+$ state of $^{12}$Be from the decay of $^{13}$Be.

\subsection{Next Generation neutron\\ detector (NGn)}

Since the construction of MoNA over twenty years ago, new techniques in photon detection as well as advanced electronics were developed offering the opportunity to design a novel neutron detector with improved position resolution and multiple neutron discrimination capability.

\begin{figure}[!b]
 \centering
\includegraphics[width=0.3\textwidth]{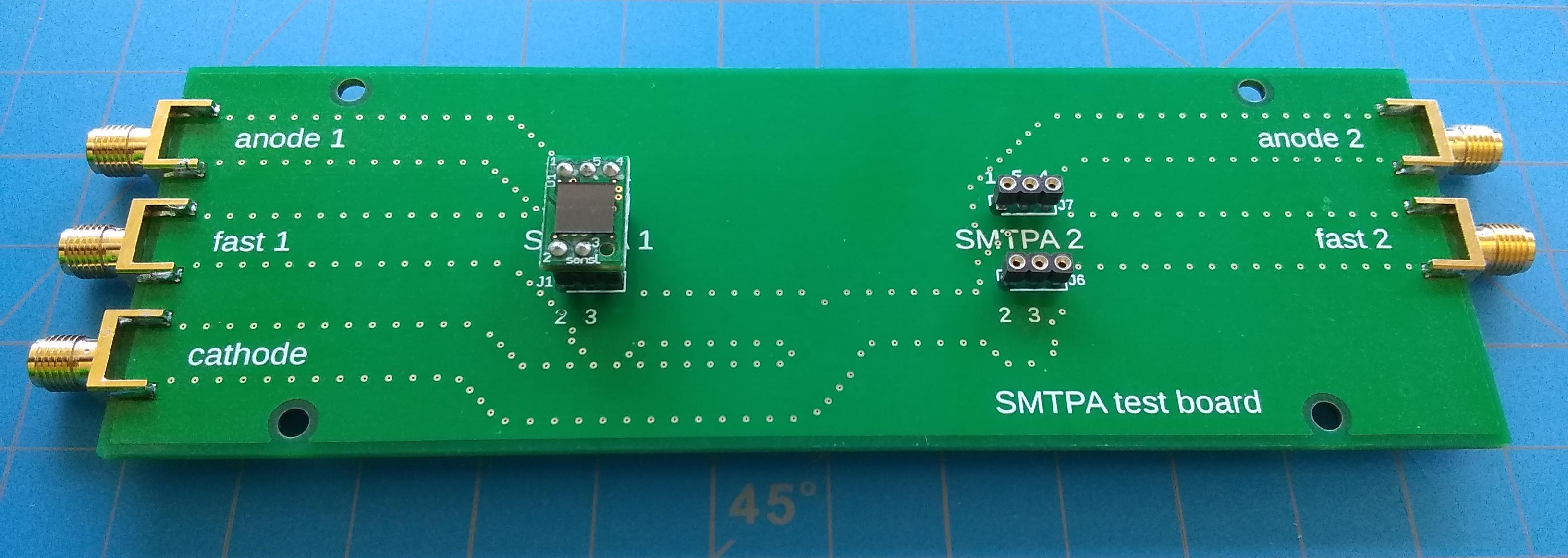}
\includegraphics[width=0.3\textwidth]{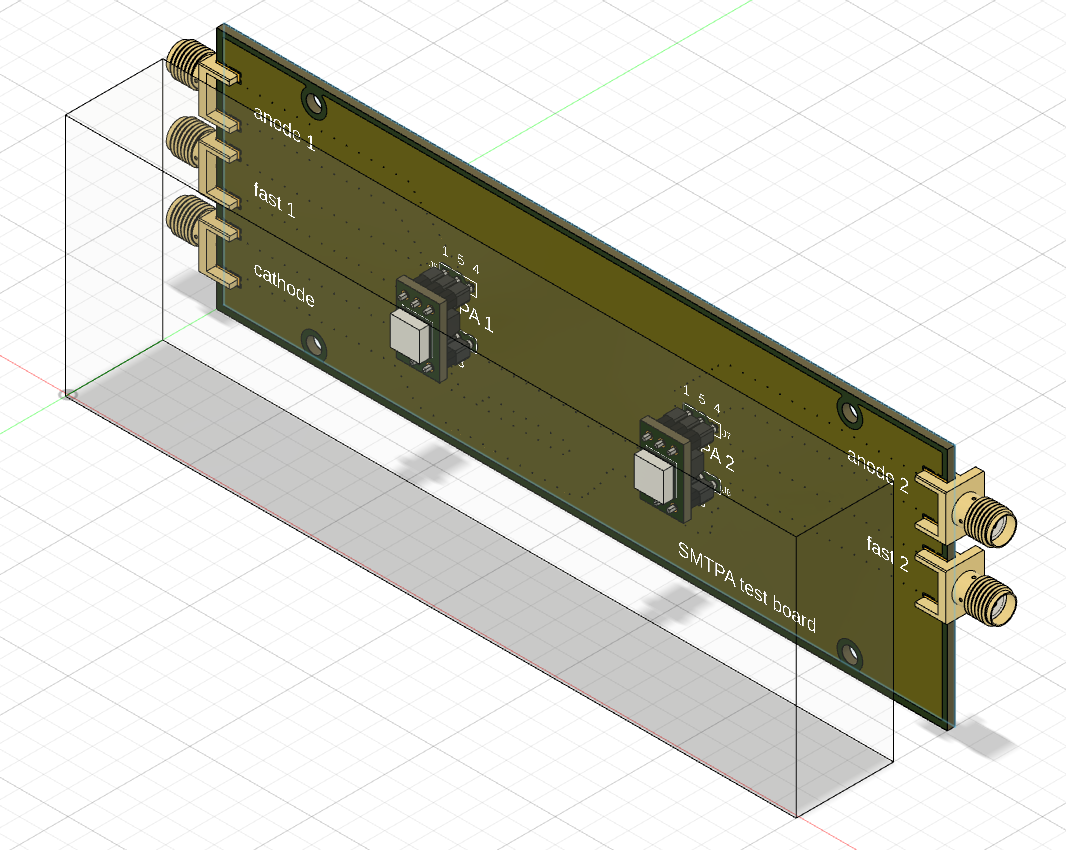}
\caption{\label{fig:smtpa}\protect  Test board to read out signals from a pair of SiPMs (one SiPM is installed, top) and assembly of small scintillation detector (bottom).}
\end{figure}

The MoNA Collaboration explored a new method to read out the light output based on an array of Silicon Photo Multiplier (SiPM) detectors. Several test kits were put together to investigate this approach and to benchmark simulation results (see figure\ref{fig:smtpa}). 

The kits were exposed to an 11-MeV mono-energetic neutron beam at TUNL where different combinations of reflective wrapping and optical coupling of the SiPMs to the plastic scintillators were tested. Also, the position sensitivity was tested by moving the detectors to different locations relative to the beam.

In 2023, the NSF funded the NGn array consisting of plastic scintillator tiles read out by SiPM sensors. The proposed detector array can be used as a stand-alone neutron detector, or in combination with MoNA-LISA to complement their detection capabilities.

\section{Educational impacts}
It has been known for a long time that undergraduate research experiences are critical to retain students pursuing science careers as well as for the education of a scientifically knowledgeable workforce.

\begin{figure*}
\vspace{-32 pt}
\centering\includegraphics[width=.87\linewidth]{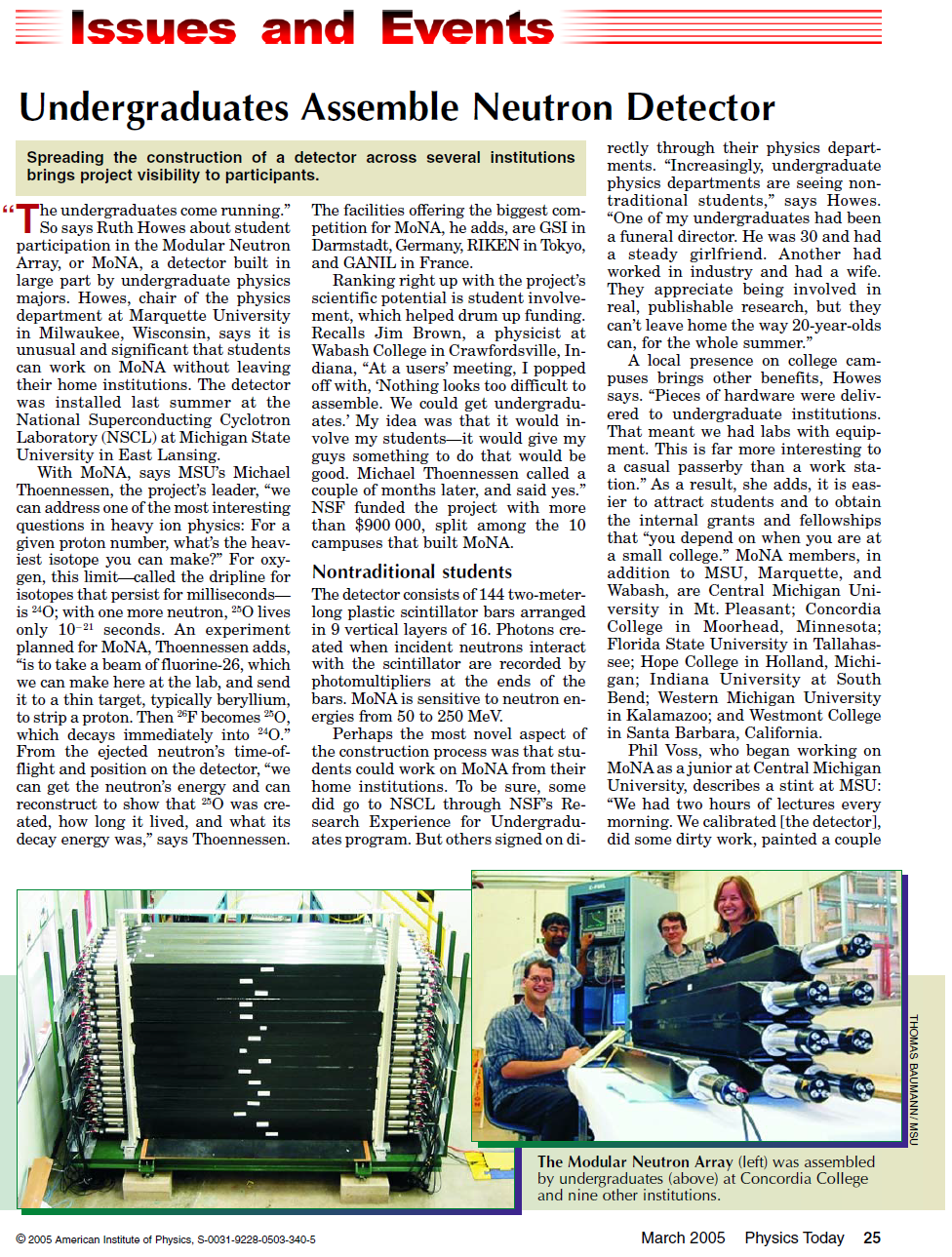}
\caption{Physics Today article describing the MoNA Project \cite{PhysicsToday}.}
\label{fig:PressRelease}
\end{figure*}

The opportunity for undergraduate research experiences was built into the MoNA Collaboration from its inception. The involvement of a group of PUIs and the modular design of the first detector array, MoNA, was an ideal setup and the basis for the success of the collaboration. The concept of the MoNA approach to research has been covered in several publications
\cite{AJP, MongolHorde, Chronicle, Symmetry, PhysicsToday, PhysicsWorld}. See, for example, the article in Physics Today shown in figure \ref{fig:PressRelease}.

\subsection{Undergraduate research\\ involvement}

\begin{figure}[!tb]
 \centering \includegraphics[width=\linewidth]{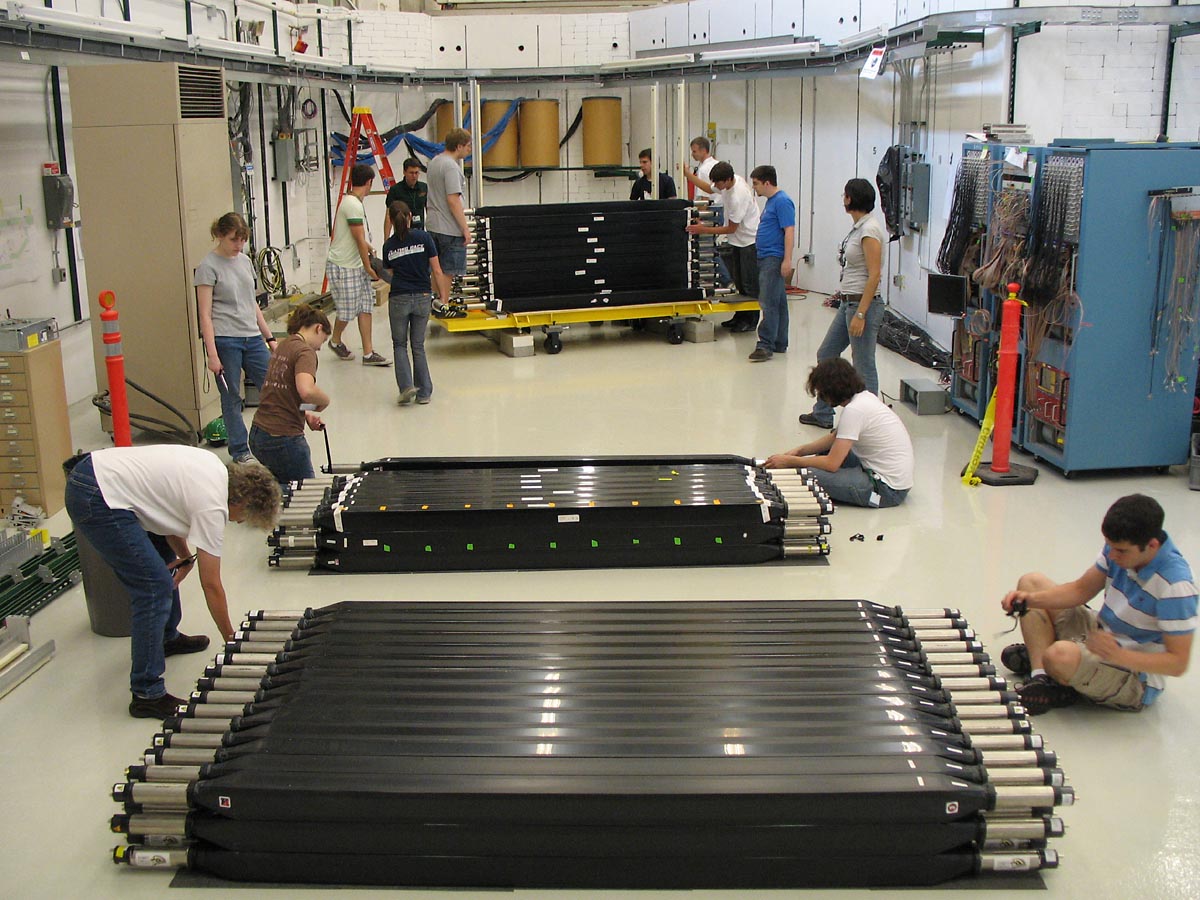}
\caption{Reconfiguration of MoNA in the N2 vault.}
\label{fig:MoNA-Move}
\end{figure}

The nine layers of the original MoNA array were assembled and tested by undergraduate students at their institutions. Once completed, they delivered the modules to the NSCL where they participated in the assembly of the entire array. This successful concept was also used for the construction of LISA. The students were also instrumental in the various moves and reconfigurations of the array. They helped with the stacking, the cabling, and setting up the related electronics (see figure \ref{fig:MoNA-Move}) each time MoNA-LISA moved for an experiment or vault location.

Once the arrays were operational, undergraduate students participated in the experiments at the NSCL. They were involved in all aspects of the experiments, from PMT gain matching, operating the detectors, running shifts, to the subsequent data analysis and physics interpretation. 

The MoNA Collaboration has found it easy to involve students in the experiments at the NSCL and now FRIB. The students can readily grasp the basic goals of the measurements. The reconstruction of the original nuclear mass is based on relativistic four-vectors. The nuclear shell model and single-particle states, while complex in detail, can easily be related to atomic shells. The students see the big picture while being involved in the experimental details. Students see moderately complex detector systems that are straightforwardly understood. (The concept of determining neutron energy from time-of-flight can be understood by first-year students.) 

Many students made sufficiently significant contributions to the overall effort to warrant co-authorship of the final publications. The 56 refereed papers of the collaboration had more than 150 undergraduate co-authors. On five papers undergraduate students were the first author, including the discovery of $^{26}$O \cite{PhysRevLett.108.142503}.

This experience within the collaboration is invaluable. The graduate students and research associates at the NSCL/FRIB provide approachable role models for them, and they feel free to ask questions of any of the faculty members in the group. For students from small undergraduate physics departments, participation in the MoNA Collaboration provides a chance to experience how physics is done in a large graduate physics department and at a world-class nuclear physics laboratory. The experience is particularly important for students who do not go on to graduate school in physics because they gain an understanding of how hard experimental scientists work to uncover the data points that underpin the theories written up in science texts and news magazines. The support of physics students who do not work as nuclear physicists but have careers in industry, K-12 education, or even the arts is important in reaching the non-scientists who set policy and control the funding for nuclear physics.

\subsection{Fostering undergraduate\\ research}

The involvement of students in more than detector assembly and running shifts required to build and evolve a research structure that allowed the undergraduates to interact and collaborate across institutions.

In 2005, communication via video conference devices started thanks to a grant from the Research Excellence Fund of Michigan to purchase digital video-conferencing equipment for all member institutions. The equipment has allowed undergraduate students to participate in the real-time acquisition and off-line analysis of data. The digital video conferencing system allowed faculty and students to have regular group, sub-group, and point-to-point meetings where pre-experiment planning was discussed and post-experimental data analysis was coordinated. The system furthered training, educating, and motivating students. 

Computational resources have evolved over the years as well. A large server was maintained at one PUI for remote access to analysis and simulation codes (mirrors of those at the NSCL) and the large data files. With the availability of cheaper high-speed computers and storage, and cloud computing solutions, data, analysis, and simulations are now shared using low-cost cloud services.

In addition to developing resources for communication and computation, the collaboration developed a culture of mentoring undergraduate students as they work on different projects. It would be difficult for single researchers from a PUI to work successfully with their students in isolation. The fact that students and faculty involved in the research participate in regular video conferences where recent results and problems can be discussed with others working on the same experiment or related analyses is crucial. This shared expertise strengthens the group effort and provides undergraduates and their faculty mentors with data analysis support.

\begin{figure}[!tb]
\centering
\includegraphics[width=0.6\linewidth]{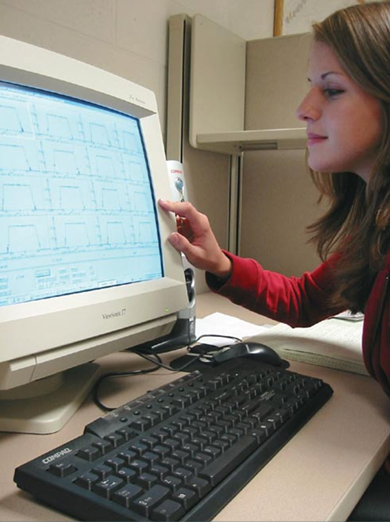}
\caption{Calibration of MoNA detectors \cite{LRP2007}.}
\label{fig:TinaPike}
\end{figure}

An undergraduate student may be responsible for analysis like a graduate student or work as part of a team to perform a portion of the analysis. Students can work with more senior researchers where they provide hours on task and have a good overview of the experiment but do not have the ultimate responsibility for the results. Undergraduate students with limited time to work can still participate by doing very focused aspects such as the calibration of a single detector subsystem, code checking, or validation of the work of others (see figure \ref{fig:TinaPike}). The same approaches are taken regarding simulation or equipment development projects.

\subsection{Collaboration structure}

In addition to performing research projects, the students also learn about the functioning and operation of a large research collaboration. Although the interaction among professors, research associates, and students are mostly informal, the collaboration established the Executive Director position. It is an elected position with a one-year term (see table \ref{tab:directors}). The Executive Director coordinates the planning and scheduling of the experiments with the NSCL/FRIB Users Office and represents the collaboration at Conferences, the Low Energy Community Meetings (LECM), and DOE/NSF Long Range planning town meetings.

\begin{table}[!b]
\caption{MoNA Collaboration Executive Directors}
\vspace*{0.1cm}
\begin{tabular}{lll}
 \hline 
Year & Director  & Institution  \\ \hline
2002	&	T. Baumann		&	Michigan State Univ.	\\
2003	&	J.E. Finck		&	Central Michigan Univ.	\\
2004	&	P.A. DeYoung		&	Hope College 		\\
2005	&	J.A. Brown		&	Wabash College 		\\
2006	&	J.D. Hinnefeld		& Ind. Univ. South Bend	\\
2007	&	W.F. Rogers		&	Westmont College 		\\
2008	&	P.A. DeYoung		&	Hope College 		\\
2009	&	B.A. Luther		&	Concordia College 		\\
2010	&	D. Meyer		&	Rhodes College 		\\
2011	&	N. Frank		&	Augustana College 		\\
2012	&	R. Haring-Kaye		&	Ohio Wesleyan	\\
2013	&	S. Stephenson		&	Gettysburg College 		\\
2014	&	W.F. Rogers		&	Westmont College 		\\
2015	&	J.A. Brown		&	Wabash College 		\\
2016	&	J.E. Finck		&	Central Michigan Univ.		\\
2017	&	P. Gueye		&	Hampton University 		\\
2018	&	S. Stephenson		&	Gettysburg College 		\\
2019	&	A.N. Kuchera		&	Davidson College 		\\
2020	&	W.F. Rogers		&	Indiana Wesleyan	\\
2021	&	N Frank		&	Augustana College 		\\
2022	&	T. Redpath		&	Virginia State Univ.		\\
2023	&	J.A. Brown		&	Wabash College 		\\
2024	&	T. Redpath		&	Virginia State Univ.		\\
 \hline
\end{tabular}
\label{tab:directors}
\end{table}

The collaboration operates under a consortium agreement, with a publication policy, a code of conduct, and a community agreement adapted from the Contributor Covenant \cite{CAgreement}.

\begin{figure}
\centering
\includegraphics[width=\linewidth]{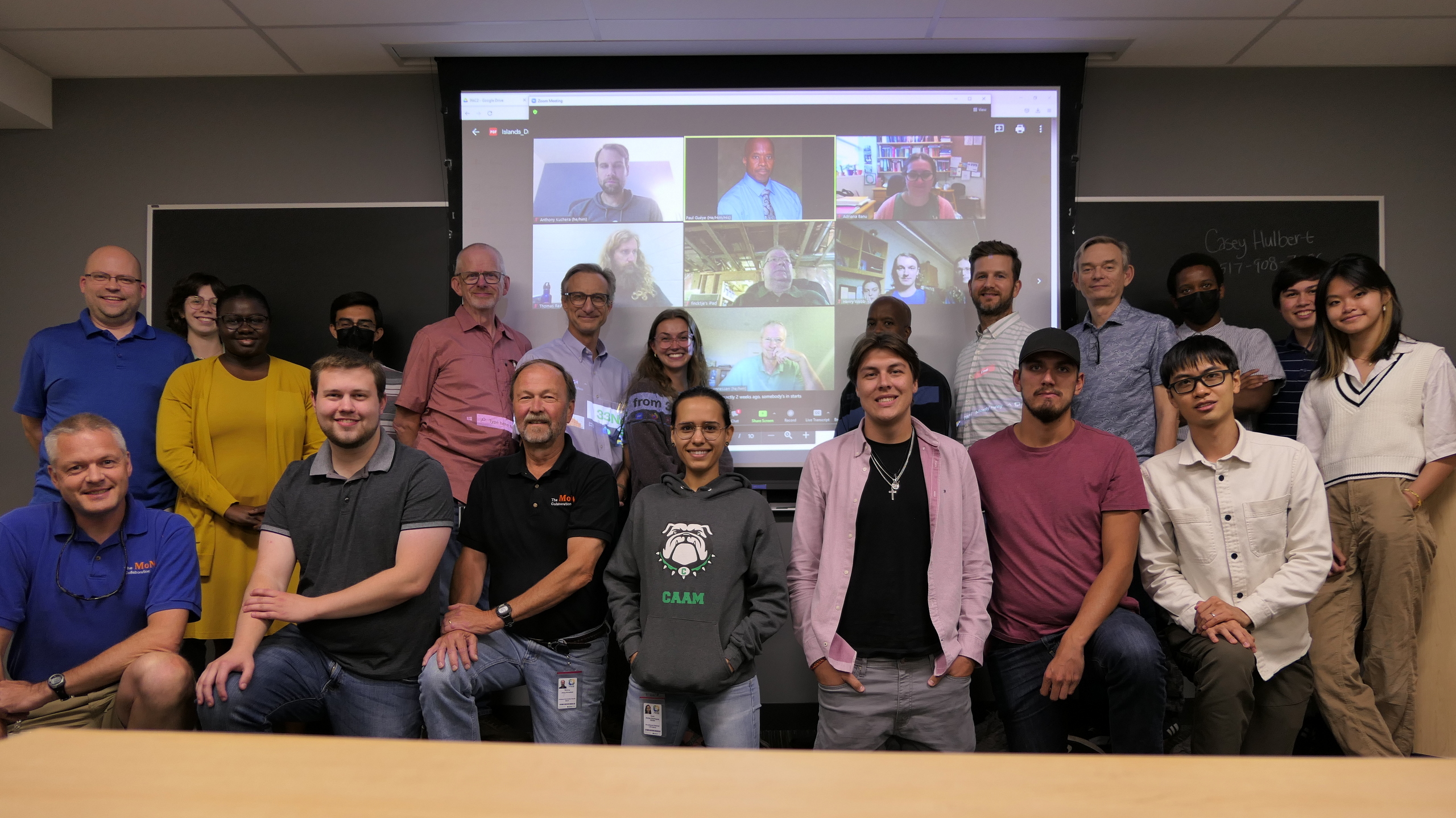}
\caption{Participants of the 2022 Collaboration Meeting at FRIB. }
\label{fig:collaborationmeeting}
\end{figure}

The students participate in all aspects of the collaboration during weekly collaboration meetings and an annual retreat. During the first ten years, the retreats were held at the Central Michigan Biological Station on Beaver Island, located near the northern tip of Lake Michigan (2004$-$2013). Subsequent retreats took place at MSU (2014, 2018, 2019, and 2022), Westmont College, CA (2015), Wabash College, IN (2016), and the MSU Kellogg Biological Station in Hickory Corners, MI (2017). The 2020 and 2021 meetings were remote only. The two most recent meetings were attached to the LECM at FRIB (2023) and the University of Tennessee at Knoxville, TN (2024). The retreats enable faculty and students to write papers, discuss analysis, develop proposals for experiments and external support, and plan for the year ahead (see figure \ref{fig:collaborationmeeting}). The annual collaboration business meeting is also held at the retreats.

\subsection{Students and research associates at MoNA}

\begin{figure}[!tb]
 \centering
 \includegraphics[width=\linewidth]{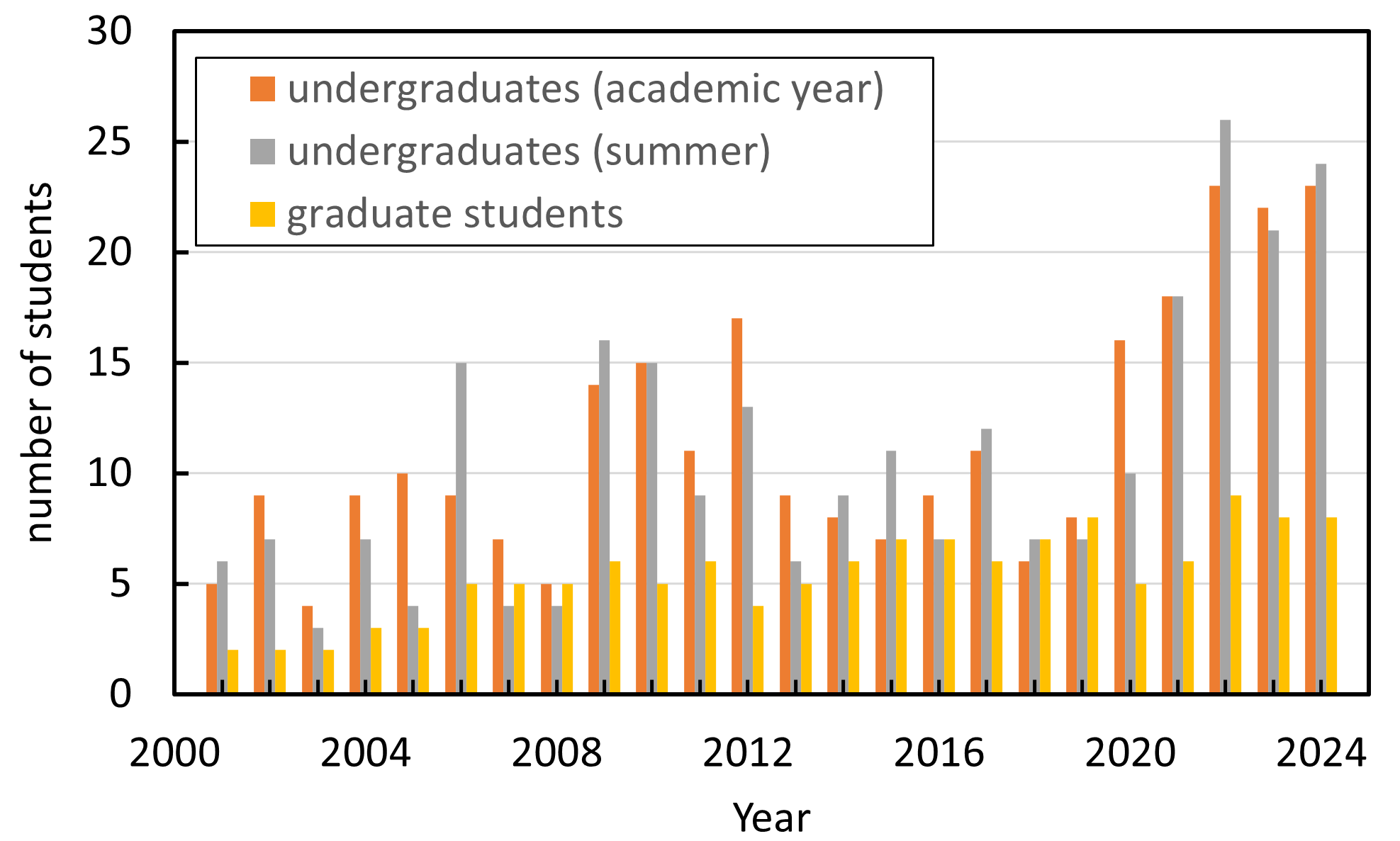}
\caption{MoNA Collaboration student involvement over the years.}
\label{fig:students}
\end{figure}

Figure \ref{fig:students} shows the number of students working with MoNA during the last 24 years. In addition to the undergraduate students enrolled at the schools of the collaboration and who worked with their professors during the full academic year, many internship students from other institutions joined the collaboration during the summers.

Overall, more than 250 undergraduate students from more than 25 different colleges and universities as well as a few high school students were involved in the research of MoNA.

The experiments at the NSCL also served as thesis projects for graduate students. Table \ref{tab:grads} lists the 16 PhD and 2 Masters theses completed with the MoNA Collaboration. Calem Hoffmann received the 2010 APS Division of Nuclear Physics (DNP) Thesis Award for his dissertation.

Research associates have been important members of the collaboration and are listed in table \ref{tab:postdocs}. They have been instrumental in developing innovative ideas, sophisticated analyses, and simulation codes. 

\begin{table*}[!tbh]
\caption{MoNA graduate students' Ph.D. theses; the two Masters thesis are indicated by a $^*$}
\begin{tabular}{lllp{9.7cm}}
 \hline 
Student & Inst.  & Year & Thesis Title \\ \hline
Nathan Frank	 & 	MSU	 & 	2006	 & 	Spectroscopy of neutron unbound states in neutron rich oxygen isotopes	\\
William Peters	 & 	MSU	 & 	2007	 & 	Study of neutron unbound states using the modular neutron array (MoNA)	\\
Greg Christian$^*$	 & 	MSU	  & 	2008	 & 	Production of Nuclei in Neutron Unbound States via Primary Fragmentation of $^{48}$Ca	\\
Calem Hoffman	 & 	FSU	 	 & 	2009	 & 	Investigation of the neutron-rich oxygen isotopes at the drip line	\\
Greg Christian	 & 	MSU	  & 	2011	 & 	Spectroscopy of neutron-unbound fluorine	\\
Michael Strongman$^*$	 & 	MSU	 	 & 	2011	 & 	Neutron spectroscopy of $^{22}$N and the disappearance of the N = 14 shell	\\
Shea Mosby	 & 	MSU	 	 & 	2011	 & 	Spectroscopy of neutron unbound states in neutron rich carbon	\\
Jesse Snyder	 & 	MSU	 	 & 	2013	 & 	Spectroscopy of $^{15}$Be	\\
Dilupama Divaratne	 & 	Ohio U. 	 & 	2013	 & 	One and two neutron removal cross sections of $^{24}$O via projectile fragmentation	\\
Rudolf Izsak	 & 	ELTE	  & 	2014	 & 	Experimental investigation of the $^8$Li  $\rightarrow ^7$Li+n Coulomb breakup process	\\
Michelle Mosby	 & 	MSU	  & 	2014	 & 	Measurement of excitation energy of neutron-rich precursor fragments	\\
Jenna Smith	 & 	MSU	  & 	2014	 & 	Unbound states in the lightest island of inversion: Neutron decay measurements of $^{11}$Li, $^{10}$Li and $^{12}$Be	\\
Michael Jones	 & 	MSU	  & 	2016	 & 	Spectroscopy of neutron unbound states in $^{24}$O and $^{23}$N	\\
Krystin Stiefel	 & 	MSU	  & 	2018	 & 	Measurement and modeling of fragments and neutrons produced from projectile fragmentation reactions	\\
Thomas Redpath	 & 	MSU	 	 & 	2019	 & 	Measuring the half-life of $^{26}$O	\\
Han Liu	 & 	MSU	 	 & 	2019	 & 	Reaction mechanism dependence of the population and decay of $^{10}$He	\\
Daniel Votaw	 & 	MSU	  & 	2019	 & 	N = 7 Shell Evolution at and Beyond the Neutron Dripline	\\
Dayah Chrisman	 & 	MSU	  & 	2022	 & 	Neutron-unbound states in the nucleus $^{31}$Ne	\\
 \hline
\end{tabular}
\label{tab:grads}
\end{table*}

\begin{table}
\centering
\caption{MoNA Collaboration research associates}
\vspace*{0.1cm}
\begin{tabular}{lll}
 \hline 
Research associate & Inst.  & Year  \\ \hline
Jean-Luc Lecouey	&	MSU	&	2003-2005 \\
Ken Yoneda	&	MSU	&	2003-2005 \\
Andreas Schiller	&	MSU	&	2003-2007 \\
Heiko Scheit	&	MSU	&	2006-2006 \\
Artemis Spyrou	&	MSU	&	2007-2009 \\
Lou Jisonna     &   Hope    & 2008-2010 \\
Zachary Kohley	&	MSU	&	2011-2012 \\
Anthony Kuchera	&	MSU	&	2014-2016 \\
Bel\'en Monteagudo Godoy	&	Hope	&	2020-2021 \\
Juan Lois Fuentes   &   MSU & 2024-present \\
 \hline
\end{tabular}
\label{tab:postdocs}
\end{table}

\subsection{Impact on undergraduates}
The benefits of engaging undergraduates in research has long been recognized by the nuclear physics community as an important goal. The MoNA model has been highlighted in the 2007 \cite{LRP2007} and the 2015 \cite{LRP2015} DOE/NSF Long Range Plans (LRP). Tina Pike from Hope College and Mustafa Rajabali from Concordia College were featured in the 2007 and 2015 plan, respectively.

\begin{figure}
 \centering
 \includegraphics[width=\linewidth]{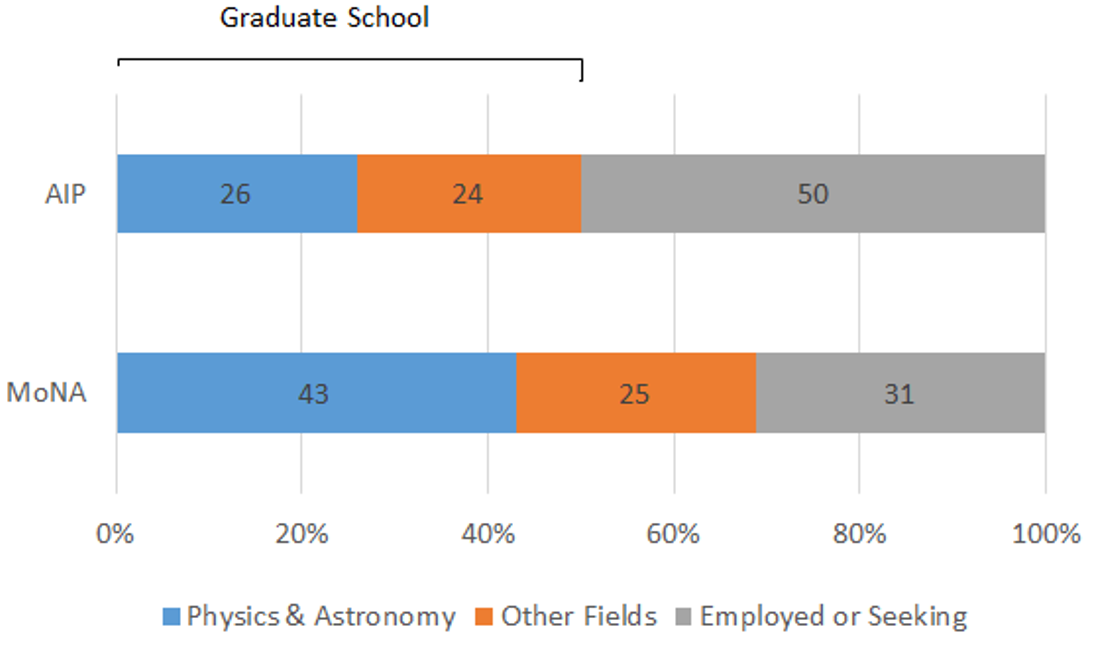}
\caption{Career choices of BS/BA graduates from bachelor's granting institutions in the U.S. from an AIP survey \cite{AIP2014} and from the MoNA Collaboration. The AIP data is from 1974 respondents from 2011 and 2012, and the MoNA data is based on 97 students from 2002 to 2014 (from \cite{Education2015}).}
\label{fig:MONAUndergradStudents}
\end{figure}

The report of the DNP town meeting on education and innovation in preparation for the 2015 plan included data on the impact on the future careers of MoNA students \cite{Education2015}: Figure \ref{fig:MONAUndergradStudents} compares data from an AIP survey \cite{AIP2014} for students with a BS/BA degree from Bachelor’s granting institutions with MoNA student data. The figure shows that a significantly larger fraction of MoNA students (69\%) chose to continue the pursuit of an advanced degree than students graduating from the average Bachelor’s granting institution (50\%). It should be noted that students from Bachelor’s granting institutions represent about one-third of all students from U.S. colleges and universities entering graduate school in physics (cited from Ref. \cite{Education2015}).  
 
\begin{figure}[!t]
 \centering
 \includegraphics[width=0.6\linewidth]{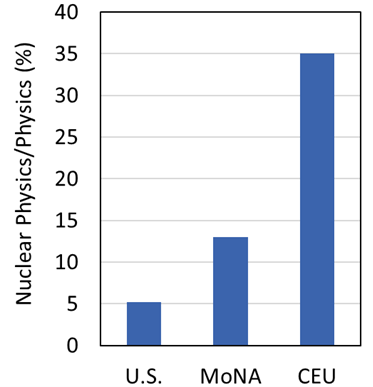}
\caption{Percentage of nuclear physics graduate students out of all physics graduate students in the U.S.\cite{Education2015}.}
\label{fig:MONAGradStudents} 
\end{figure}

The report also demonstrated that the percentage of MoNA students selecting nuclear physics in graduate school is larger than the U.S. fraction of the average number of PhDs from 2000$-$2012. The fraction is even larger for students who participated in the DNP Conference Experience for Undergraduate (CEU) students (see figure \ref{fig:MONAGradStudents}). It should be mentioned that the MoNA students are strongly encouraged to apply for the CEU program and on average about five students get selected and participate in the program each year.

\begin{figure}[!h]
 \centering
 \includegraphics[width=\linewidth]{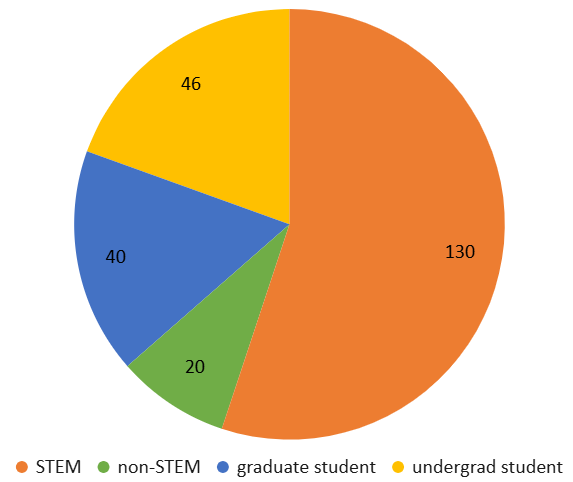}
\caption{Current job distribution of MoNA students (81.7\% reported).}
\label{fig:MONAStudentsJobs}
\end{figure}

The MoNA Collaboration continues to track the careers of the students. Figure \ref{fig:MONAStudentsJobs} shows the current job distribution of former MoNA students. About 75\% of students go into graduate school or are pursuing a STEM career illustrating the significant impact of MoNA on the STEM workforce.

\section{Future plans}

The transition of the NSCL to FRIB has been completed and FRIB began operation in 2022. As part of the reconfiguration, the MoNA-Sweeper setup is being relocated to the S2 vault. The MoNA Collaboration has already four approved experiments to be performed in the new location \cite{PAC}.

\begin{figure}[!tb]
 \centering \includegraphics[width=0.7\linewidth]{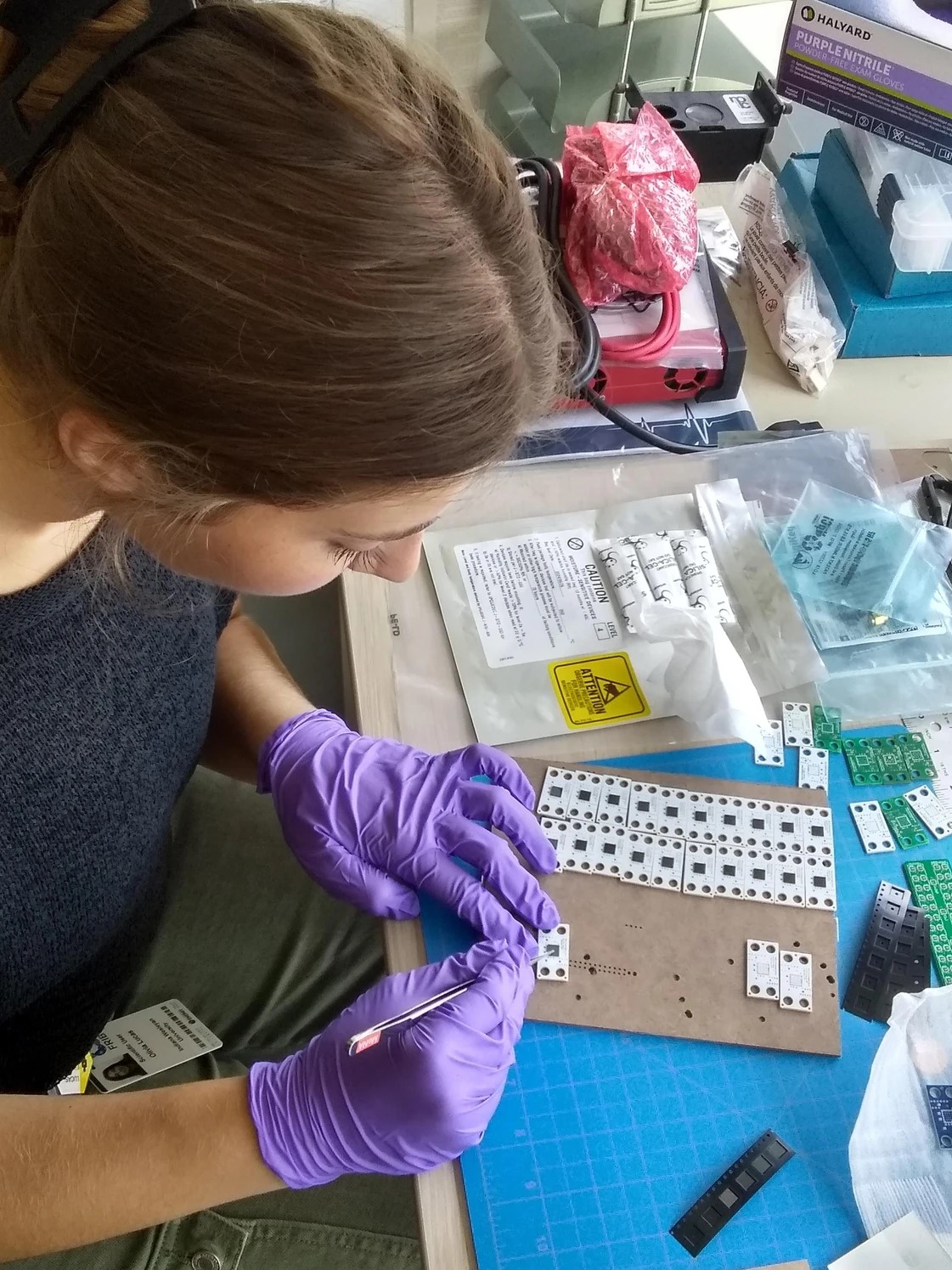}
\caption{Assembly of a NGn prototype.}
\label{fig:NgN-Assembly}
\end{figure}

The two experiments approved by the first Program Advisory Committee (PAC)  are the observation of neutron-unbound $^{30}$F ($\#$21016) and the study of unbound excited states in $^{53}$Ca ($\#$21066). The second PAC approved the investigation of the halo structure of $^{37}$Mg ($\#$23033) and the study of a possible $p$-wave halo in the $^{34}$Na ground state ($\#$23068). The first experiments are expected to run in 2025.

The MoNA Collaboration also continues with the development of NGn which will be completed by 2026. The project continues the MoNA tradition of involving undergraduate students in the research (see figure \ref{fig:NgN-Assembly}). Each participating institution received a test kit and the testing results are shared in weekly video conference meetings. The undergraduates will also have the opportunity to participate in future module testing at TUNL and LANSCE.

In the longer term future, MoNA will be relocated to the High Rigidity Spectrometer (HRS) which is currently under construction \cite{HRS,NOJI2023167548}.

\section{Summary}

Cutting-edge physics remains to be done utilizing fast fragmentation beams. The evolution of shell closures (magic numbers) as the stabilizing influence of protons in the same orbitals is lost for the most neutron-rich nuclei, which continues to be of particular interest. An additional focus is the study of neutron pairing correlations, which can be studied using neutron-rich nuclei in which sequential two-neutron decay is energetically forbidden, and only direct two-neutron decay can occur. Moreover, reaction studies and cross-section measurements can reveal, e.g., neutron and radiative strength functions. Reactions on exotic nuclei involving neutrons are also often of importance for explosive scenarios (such as r-process) in astrophysics.

The MoNA Collaboration has been able to take advantage of the areas of expertise of its members to create a collaboration that has effectively involved undergraduate students from its beginning and continues to do so to this day. Students readily understand the nature of these experiments and can participate in meaningful ways. The impact on students by exposing them to the international-level research currently conducted at FRIB is significant and helps to train the next generation of physicists.

Based on the success of the past 25 years, the MoNA Collaboration is poised to play an important role in educating the next generation of nuclear physicists, discovering new physics at the edge of existence, and developing new devices for future experiments at FRIB.

\newpage

\section*{Appendix: Previous director's \\ statements}

The MoNA Collaboration consists of a group of researchers, most from primarily undergraduate institutions, who are pursuing studies of nuclei close to and beyond the neutron dripline using the Modular Neutron Array (MoNA). These experiments can only be done with neutron-rich nuclei produced via projectile fragmentation, as carried out, for example, at the National Superconducting Cyclotron Laboratory (NSCL) at Michigan State University, where MoNA is currently located.

Since the first detectors of MoNA arrived for assembly in 2002, 64 undergraduate and high school students (as of Spring 2007) have participated in cutting-edge research in nuclear physics as part of the MoNA Collaboration. These students have assembled and tested the components of MoNA, participated in MoNA experiments and workshops at the NSCL and in the annual collaboration retreat, and played a central role in data analysis.

The MoNA Collaboration has been a model for involvement of undergraduates in forefront research. The collaboration is committed to continuing its role in the study of nuclei at the limits of stability and in the training of the next generation of nuclear scientists. Our experience over the last six years leads us to the following observations:
\vspace*{-0.35cm}

\begin{itemize}
\item
    Studies of nuclei at the neutron dripline utilizing beams produced by fast fragmentation produce cutting-edge science. These experiments are well suited to meaningful participation by undergraduate students in a multi-institution collaboration.
    \vspace*{-0.3cm}
\item
    The collaboration has thrived in a university setting, where undergraduate education is at the core of the institutional mission.
\end{itemize}
\vspace*{-0.35cm}

We look forward to a next-generation facility for rare-isotope beams which would ensure the continuation of this successful scientific and educational collaboration for years to come.\\
\vspace*{-0.1cm}

\noindent Jerry Hinnefeld\\
\emph{Executive Director, the MoNA Collaboration}\\
{\small South Bend, January 17, 2007}

Since the last version of this document the MoNA Collaboration has continued to thrive and grow.  More than 100 undergraduate students have now been part of the collaboration's scientific endeavors
playing vital roles in the study of the nuclei at the limits of stability. Our collaboration has grown in other ways as well.  New institutions and investigators have joined the collaboration. Sharon
L. Stephenson (Gettysburg College), Nathan Frank (Augustana College in Rock Island, IL), Artemis Spyrou (Michigan State University),
Robert A. Kaye (Ohio Wesleyan University) and Deseree Meyer Brittingham (Rhodes College) are bringing new skills and insights to
the collaboration's work.  In addition a new detector system is under construction by undergraduates at the collaboration schools.
LISA, the Large multi-Institution Scintillator Array, will work in conjunction with MoNA to increase our ability to measure angular distributions of reaction neutrons as well as improve the resolution and efficiency of detection in our experiments.

The MoNA Collaboration has always been forward-looking whether in the preparation of the next generation of physicists or in the construction of detectors that are ready for use in the next generation of rare isotope beam facilities (FRIB).  Today, we see a bright future for the collaboration, the NSCL, and rare-isotope physics.\\

\noindent Bryan A. Luther \\
\emph{Executive Director of the MoNA Collaboration}\\
{\small Moorhead, MN,  Sept. 9, 2010}

\vspace*{8ex}
In the last two years, the MoNA Collaboration has completed LISA, the Large multi-Institutional Scintillator Array. Twenty-three undergraduates worked on construction, testing, and installation of LISA, with additional students playing key roles in data analysis. A successful commissioning experiment in June 2011 continues our scientific program of probing nuclei at the limits of stability. The higher efficiency and better resolution of MoNA LISA combined will allow the collaboration to study a wide array of isotopes that will be available when the Facility for Rare Ion Beams (FRIB) comes online. Extensive, meaningful undergraduate involvement in the cutting-edge science provides pivotal research experiences for students and contributes to training the next generation of nuclear scientists. The collaboration continues to exemplify a successful partnership between primarily undergraduate institutions and a large research university. We are excited about future research and educational opportunities that will be possible with FRIB and as our collaboration continues to grow.\\

\noindent Deseree Meyer Brittingham\\
\emph{Executive Director, the MoNA Collaboration}\\
{\small Beaver Island, MI, August 20, 2011}\\

\vspace*{8ex}
The MoNA Collaboration has continued to demonstrate growth in its scientific and educational objectives and outcomes since the production of the last White Paper.  Since the beginning of 2012, 15 papers in refereed journals were published collectively by the collaboration, including cutting-edge studies of the ground-state dineutron decay of $^{16}$Be and two-neutron radioactivity in the decay of $^{26}$O.  A hodoscope particle detector array, intended to increase the sensitivity of the identification of charged fragments, was developed by Augustana College and was implemented in a commissioning experiment at the NSCL last summer.  Paul Gueye (Hampton University) has joined the collaboration and is involved in an effort to develop a segmented target, which will determine the location of nuclear reactions within the reaction target and thus provide better resolution in decay energy measurements.  Additionally, our mission to help educate the next generation of scientists remains an important cornerstone of our work.  Two NSCL graduate students received their Ph.D. in MoNA-related research and over 20 undergraduates from across the participating institutions of the collaboration were involved in research projects in 2012--2013.  We also continue to keep a keen eye to the future, making preparations for our experimental program to be a possible "Day One" user of the new Facility for Rare Isotope Beams (FRIB), currently slated for completion in 2022.\\

\noindent Robert Kaye\\
\emph{Executive Director, the MoNA Collaboration}\\
{\small Beaver Island, MI, August 17, 2013}

\vspace*{8ex}
The MoNA (Modular Neutron Array) Collaboration continued to find success over the past year. To date, we have 37 peer reviewed papers with over half of those having undergraduate students as co-authors.  In 2014 three graduate students completed their PhDs and another has data in hand to study the energy gap between the sd--pf neutron shells in $^{25}O$.  This year the total number of MoNA Collaboration undergraduate students has surpassed our lucky number of 144 -- the number of neutron detectors in MoNA or LISA.  Our 147 undergraduate students have presented over fifty times at national physics conferences. The infrastructure of the MoNA Collaboration and the tradition of expecting quality work from our students at all levels of their academic careers has led to our improving research opportunities and preparing the next generation of physicists.\\

\noindent Sharon Stephenson\\
\emph{Executive Director, the MoNA Collaboration}\\
{\small East Lansing, MI July 20, 2014}\\

\vspace*{8ex}
At the time of this 2016 MoNA (Modular Neutron Array) Report, our collaboration remains as strong as ever.  Since we first began 13 years ago using MoNA for nuclear physics experiments, we have published 40 peer reviewed articles, primarily in Physical Review Letters, Physical Review C, Physics Letters B, Nuclear Physics A, and Nuclear Instruments and Methods A, with over half of them including undergraduate co-authors.  Recent scientific highlights of our group's work include a study of neutron correlations in the decay of excited $^{11}$Li, selective population of unbound states in $^{10}$Li, population of $^{13}$Be using charge-exchange reactions, characterization of low-lying states in $^{12}$Be, a search for unbound $^{15}$Be states using the $^{12}$Be + 3n channel, three-body correlations in the decay of the $^{26}$O ground state, analysis of $^{10}$He production mechanism using a $^{14}$Be secondary beam and a deuterated target, and a measurement of the low-lying excited states of $^{24}$O (which served as the LISA commissioning run). Recently completed experiments under analysis include a study of the of the equation of state using rare isotope beams, knockout reactions on p-shell nuclei, and in summer 2015 an experiment to measure the ground state energy of $^{10}$He using two separate production mechanisms was completed. Approved experiments for the near future include a measurement of the $^{9}$He ground state, and lifetime measurements with a decay-in-target method.  To date 8 graduate students have completed their PhD degrees in MoNA research, 2 will be completing them soon, and 3 are relatively new to the group.  By now 159 undergraduate students have participated in MoNA research, and have presented their research 56 times at national physics conferences. And for the first time our summer working retreat workshop was held outside of the state of Michigan, in sunny Santa Barbara, CA.  The main goal of our collaborative effort is the execution of high quality research in nuclear science, with undergraduate participation at the heart of our efforts.  This vision will continue to drive our efforts in future years with a shared goal of helping inspire and prepare the next generation of physicists.\\

\noindent Warren Rogers\\
\emph{Executive Director, the MoNA Collaboration}\\
{\small Santa Barbara, CA, July 31, 2015}\\

\vspace*{8ex}
The MoNA Collaboration had another successful year and has been busily preparing for the completion of FRIB and the eventual construction of the High Rigidity Spectrograph. The addition of our active target system allows for experiments with even more exotic nuclei, and the system saw its first use in an experiment to measure the groundstate lifetime of $^{26}$O. The collaboration continues to do cutting edge science in the structure and reactions of the most neutron-rich nuclei accessible. Beyond the scientific and technical the project continues to shape the careers of students, post-doctoral researchers, and faculty alike. The collaboration has touched the lives and careers of 171 undergraduates, who have participated in the running and analysis of our experiments or the construction of the detectors. This work has resulted in 44 peer-reviewed publications and many conference presentations, proceedings, and CEU posters. The collaboration remains vibrant and effective.\\

\noindent James Brown\\
\emph{Executive Director, the MoNA Collaboration}\\
{\small Crawfordsville, IN, August 13, 2016}\\

\vspace*{8ex}
The MoNA Collaboration had another successful year of science and education, and we have also been presented with future challenging technical and personnel changes. In July the entire Collaboration, including 12 undergraduates, conducted an experiment to measure the $^9$He ground and excited states. We are now busy preparing in late November to look for neutron unbound states in the island of inversion.  This will possibly be our last experiment in the N2 vault.  To permit the testing of a CycStopper, we were asked to move the MoNA LISA detectors and Sweeper.  In order to continue our scientific program we developed a plan to place our devices in front of the S800.  We submitted a Letter of Intent to PAC 41 and they recognized and agreed with our proposal that this ``would enable interesting studies on the nuclear structure and reactions involving the population of neutron-unbound excited states in medium mass neutron-rich nuclei, by giving improved PID resolution necessary to identify higher mass fragments.''  To stimulate interest in an experimental MoNA LISA/Sweeper/S800 campaign, the Collaboration led a Working Group at the 2017 Low Energy Community Meeting where six potential experiments were presented and discussed.  We will be busy leading up to PAC 42 in March of 2018 writing proposals and addressing the technical issues for this experimental campaign.

Sadly, in mid-June Michael Thoennessen informed us that the APS Board of Directors approved his appointment as the new APS Editor in Chief. This is a great honor and opportunity for Michael.  We are proud of him.  But we are very sad to see him go.  Without Michael there would be no MoNA Collaboration and nearly 200 undergraduate students would have missed an unparalleled scientific opportunity.  While we are indebted to him for his scientific leadership, we will mostly miss him as a friend.\\

\noindent Joe Finck\\
\emph{Executive Director, the MoNA Collaboration}\\
{\small Gull Lake, MI, August 12, 2017}\\

\vspace*{8ex}
The National Superconducting Cyclotron Laboratory is in its last years of operation and the construction of its upgrade, the Facility for Rare Isotope Beams, is near completion with an expected start date around 2021/2022. The MoNA Collaboration has established itself as one of the most successful collaboration with unprecedented impact in undergraduate physics training (more than 200) through its decade of existence. Over the past year, the Collaboration achieved several milestones toward its transition from NSCL to FRIB science. One experiment to study neutron unbound states in the island of inversion was completed in the Fall 2017 and another experiment was approved by the PAC42 centered on the MoNA/LISA neutron detector arrays. A NSF/MRI to build a Si/CsI based telescope was awarded (N. Frank, PI) to improve the identification of heavy fragments. The expertise of the MoNA Collaboration also started expanding its impact on non-MoNA science at NSCL by participating in the SUN experiment over the Summer 2018 (P. DeYoung). The MoNA/LISA campaign in the N2 vault has ended after 14 years of operation. The entire experimental setup is being moved to the S2 vault for a future exciting and productive research program. The Collaboration contributed to the 2018 Nuclear Structure and Low Energy Community Conferences over the Summer with posters and oral presentations highlighting its research and impact in the field of fast neutrons science, and several students and faculty will be attending and presenting at the 2018 DNP meeting in the Fall. FRIB has initiated the investigation of an upgrade from 200 MeV/u to 400 MeV/u and the Collaboration is making plans for new detectors. P. Gueye has accepted a new position with FRIB/MSU. Through this new appointment, the MoNA Collaboration is reorganizing itself to grow its science for a bright future.\\

\noindent Paul Gueye\\
\emph{Executive Director, the MoNA Collaboration}\\
\small{NSCL/FRIB, East Lansing, August 12, 2018}\\

\vspace*{8ex}
Summer of 2019 finds the MoNA Collaboration looking back at a successful year and looking forward to a variety of challenges and opportunities.  A number of our Ph.D. students are completing their degrees and starting the next chapter in their careers.  Our undergraduate students are making meaningful contributions to our publications and well-positioned for graduate programs and STEM fields.  Faculty have reached milestones, changed home institutions, and been awarded federal funding for our work.  Over the past year we have dealt with physical changes – a large-scale move from one experimental area to another, detector development and data acquisition upgrades, as well as planning for new experiments -- two at the NSCL/FRIB and one at Los Alamos National Laboratory.  Our productivity is tied to our group’s commitment to educating the next generation of scientists while pursuing new and exciting physics.  We anticipate an exciting year ahead! \\

\noindent
Sharon Stephenson\\
\emph{Executive Director, the MoNA Collaboration}\\
{\small NSCL, July 19, 2019}\\

\vspace*{8ex}
The year 2020 will be remembered historically as the year of COVID-19. The low energy nuclear physics community will also remember it as the end of the NSCL era. The final two MoNA experiments at the NSCL were scheduled for summer 2020 but due to the pandemic were delayed. The Collaboration anticipates running at least one of the two approved experiments before the NSCL shuts down in an unfortunately shortened run schedule. As one era ends another begins. The MoNA Collaboration continues to plan and develop looking ahead to the FRIB era. This has been done through the development of next generation detectors, simulations, analysis techniques, and a deeper understanding of the MoNA bars. The Collaboration successfully ran a second experiment at LANSCE Fall 2019 and a test run at NSCL in Winter 2020 to commission the newly developed charged particle telescope. Three MoNA graduate students have defended their dissertations and have begun new jobs. Several undergraduate students have been involved and made an impact with many participating in the 2019 CEU program at the fall DNP meeting. There is a lot of uncertainty in the world right now, but the MoNA Collaboration has always found a way to rise to the occasion. This next year will be no exception.\\

 \noindent
Anthony Kuchera\\
\emph{Executive Director, the MoNA Collaboration}\\
{\small Davidson College by ZOOM, July 30, 2020}\\

\vspace*{8ex}
During this interim period between the NSCL shutdown and the start of full FRIB operations, and while the country and the world continue to struggle with the COVID-19 virus, the MoNA Collaboration has remained remarkably busy and productive. The collaboration conducted its final NSCL-based experiment in September 2020, and submitted three experiment proposals for the first round of experiments at FRIB.  It has also been busy in active and polarized target design and construction, charged fragment telescope development and implementation for use in experiments without the Sweeper Magnet, DDAS development and implementation, Monte-Carlo simulation development, analysis of data from previous experiments for physics involving other isotope-neutron correlations, analysis of neutron scattering data from experiments at LANL, development of multi-neutron sorting algorithms and machine learning for datasets involving multi-neutron decays, and development of next generation neutron detectors and array designs for use at FRIB.  Student and post-doctoral participation in MoNA remains strong, evidenced in part by the number of 2021 MoNA Collaboration meeting presentations given by 6 high school, 11 undergraduate, and 7 graduate students, as well as by 3 post-doctoral associates.  The future of the MoNA Collaboration remains bright, ensured in large part by the significant quality and talent of our younger PIs, who as a group will help carry MoNA into the future.  And as I pass executive leadership for this coming year to Nathan Frank, I am pleased to announce the addition of our newest PIs, Thomas Redpath and Calem Hoffmann.  Despite the challenges of these past two years, the MoNA Collaboration remains strong.\\

 \noindent
Warren Rogers\\
\emph{Executive Director, the MoNA Collaboration}\\
{\small Indiana Wesleyan University by ZOOM, August 2, 2021}\\

\vspace*{8ex}
FRIB officially started operations in 2022 and the MoNA Collaboration is actively preparing for the future. The collaboration has spent the year preparing to run two experiments, $^{30}$F and $^{53}$Ca, which were accepted by PAC1 at FRIB and likely will run in early 2024. Working groups were formed along topics of Experiment Infrastructure, Detector Systems, Data Acquisition, Simulation and Analysis Software, PAC1 Project Planning, and Electronic Communication and Documentation, to make sure that we are ready for these experiments. In addition to future preparations, the collaboration worked on an analysis of $^{13}$Be data using a sweeper-less setup, analysis of data from previous experiments for physics involving other isotope-neutron correlations, analysis of new neutron scattering data from experiments at LANL, development of multi-neutron sorting algorithms and machine learning for datasets involving multi-neutron decays. The two-day annual MoNA Collaboration retreat included Primary Investigators (PIs), graduate students, undergraduate students, and high school students that contributed to the MoNA research program over the last calendar year presenting on the topics listed above along with a discussion of PAC2 proposal ideas and the next generation neutron detector among other new experimental devices. We welcomed a new PI Adriana Banu (James Madison University) as a full member of the Collaboration as one of eight collaborators on the Next Generation Neutron Detector and welcomed back PI Michael Thoennessen. With the growth of the collaboration and hard work of this year, the next Executive Director Thomas Redpath will ensure that we are successful in preparations for the first experiments at FRIB.\\

 \noindent
Nathan Frank\\
\emph{Executive Director, the MoNA Collaboration}\\
{\small FRIB, East Lansing MI, August 15--16, 2022}

\vspace*{8ex}
Reflecting on yet another busy and productive year for the Collaboration, one highlight was the submission of an MRI proposal to build a next-generation neutron detector to expand the
Collaboration’s capacity for studying neutron-unbound systems. Led and written by Thomas Baumann, this milestone comes on the twentieth anniversary of the first MoNA detectors arriving for assembly. That construction effort was largely carried out by students and PIs at primarily undergraduate
institutions. Twenty years later, more than 250 undergraduate students have worked on MoNA-related projects, and the Collaboration continues to add to that number as the MRI proposal, if funded, will again involve the efforts of undergraduate institutions in constructing the new detectors.

The Collaboration was fortunate to welcome two new PIs: Belen Monteagudo from Hope College joined in August 2022 and Aldric Revel, MSU/FRIB, joined in July 2023.
The second FRIB PAC approved two MoNA experiments to use Coulomb breakup reactions to probe the neutron configurations, separation energies, and geometrical information for $^{34}$Na (a
possible neutron halo nucleus) and $^{37}$Mg (evidenced to be a neutron halo nucleus). This is an impressive feat given the competitive selection process – only 35\% of the 11,859 requested facility-use hours were approved. Congratulations to Belen and Aldric for their work preparing the
proposals. This brings the total number of approved FRIB-era MoNA experiments to four.

The Collaboration continues to work towards running its first experiments at FRIB. Tasks include redesigning the Sweeper drift chambers and time-of-flight detectors to improve particle identification
capabilities. Meanwhile analysis efforts continued for the last MoNA experiment at NSCL and the 2019 and 2022 neutron scattering measurements at LANSCE. July 2023 saw a trip to LANSCE to prepare for another measurement using diamond detectors to study neutron scattering on carbon. All of these activities will ensure a busy and exciting year for this Collaboration of outstanding scientists.\\

 \noindent
Thomas Redpath, Virginia State University\\
\emph{Executive Director, the MoNA Collaboration}\\
{\small FRIB, East Lansing MI, August 7--8, 2023}

\vspace*{8ex}

The year's highlight was NSF's awarding funding for our MRI proposal to build a next-generation neutron detector! Over the summer (and before), students got busy assembling prototypes, working on simulations, assessing various approaches to data analysis and position reconstruction, and five of the PIs and their students traveled to the Triangle Universities Nuclear Laboratory (TUNL) at Duke University to test various prototypes.  The group included five undergraduate and two graduate students, for some of whom this was their first research trip. Special thanks to Anthony Kuchera for making this connection with TUNL. 

The collaboration was well represented at the fire-delayed 2023 DNP with a total of twenty oral and poster presentations. Special thanks to Hawaii resident and MoNA PI, Bryan Luther, who helped to house and care for persons displaced by those tragic fires. In the spring, the collaboration again showed a strong presence at the APS April meeting with another dozen presentations. This strong showing highlights the continued efforts of the PIs, their students.

The collaboration is strong as we enter our twenty-fifth year!\\

\noindent James Brown, Wabash College\\
\emph{Executive Director, the MoNA Collaboration}\\
{\small Crawfordsville, IN, October, 2024}\\

\newpage
\bibliography{MoNA25}

\end{document}